\documentclass[12pt]{article}
\usepackage{amsmath,amssymb,graphicx}
\usepackage{geometry}

\usepackage{amsmath,amssymb,graphicx}
\usepackage{booktabs}
\usepackage{array}
\usepackage{geometry}
\usepackage{subcaption}
\usepackage{caption}
\usepackage{float}
\usepackage{lineno}
\modulolinenumbers[1]

\usepackage{setspace}
\let\oldthebibliography\thebibliography
\renewcommand{\thebibliography}[1]{%
	\oldthebibliography{#1}%
	\setlength{\itemsep}{0pt}%
	\setlength{\parskip}{0pt}%
}
\title{
	\textbf{Analytical Modeling of Thermal Shield-Induced Tilt Noise in Lunar Regolith: Constraints on Future Lunar Seismometers and Gravitational waves Detectors}
}

\author{
	\Large
	Sreejaya Kizhaekke Pakkathillam$^{1}$\thanks{Currently at: California Institute of Technology, Pasadena, CA, USA},
	Aneesha Ann Joseph$^{1}$, \\
	\Large Philippe Lognonné$^{1}$ \\[0.3cm]
	\normalsize
	$^{1}$Université Paris Cité, Institut de Physique du Globe de Paris, CNRS, France.
}
\date{} 

\begin{document}
	\maketitle
	\clearpage
	
	\begin{abstract}
		Surface features such as thermal shields, insulating blankets, and exposed rocks modify the absorption of solar radiation in the lunar regolith, producing lateral temperature gradients missed by one-dimensional thermal models. These gradients cause thermoelastic deformation that tilts surface-deployed seismic instruments, adding long-period noise. We develop an analytical framework, based on Hankel transforms in a semi-infinite regolith, that treats these features as perturbations to a laterally uniform diurnal thermal state; linearity allows multiple features to be superposed. The framework is applied to the Apollo~14 Passive Seismic Experiment, including the seismometer thermal shield and surrounding instruments. The Apollo shield is a multilayer insulation (MLI) blanket that passes a fraction of the solar-flux fluctuation to the ground. Near local noon the shadow of the instrument body shifts by 2-5~cm across this dimly lit ground. This asymmetry dominates the tilt; its median amplitude spectral density reaches $\approx7$-$8\times10^{-11}\,\mathrm{rad}/\sqrt{\mathrm{Hz}}$ in the 1-5~mHz band, about three orders of magnitude above that of an opaque shield. Enlarging an opaque shield lowers the tilt, but the body-shadow tilt of an MLI shield does not decrease with shield radius. Thermal heterogeneity can thus generate a tilt bias, present even during seismically quiet intervals, that should be included in the noise budget of future lunar horizontal seismometers and inertial GW accelerometers, and the optical transmission of the shield and the instrument's own shadow can matter more than shield size. The analytical framework provides a computationally efficient way to evaluate thermal-shield geometry, albedo contrast, and instrument placement for future lunar seismic and inertial gravitational-wave sensor deployments.
	\end{abstract}

	\section*{Glossary of Symbols}
	\begin{tabular}{p{0.2\linewidth}p{0.74\linewidth}}
		$(r,\theta,z)$ & Cylindrical coordinates: radial distance [m], azimuth [rad], depth (positive downward) [m] \\
		$R_s$, $R_j$ & Radius of the thermal shield and of surface feature $j$ [m] \\
		$d_{\ell j}$ & Distance between seismometer leg $\ell$ and feature $j$ [m] \\
		$\vartheta$ & Temperature perturbation [K] \\
		$T_0$ & Mean equilibrium surface temperature [K] \\
		$\rho$ & Density of the regolith [kg m$^{-3}$] \\
		$C_p$ & Specific heat capacity [J kg$^{-1}$ K$^{-1}$] \\
		$\kappa$ & Thermal conductivity [W m$^{-1}$ K$^{-1}$] \\
		$\chi$ & Thermal diffusivity, $\kappa/(\rho C_p)$ [m$^2$ s$^{-1}$] \\
		$\delta$ & Thermal skin depth [m] \\
		$\sigma$ & Stefan--Boltzmann constant [$5.67 \times 10^{-8}$ W m$^{-2}$ K$^{-4}$] \\
		$A$, $A_R$ & Albedo; albedo of the regolith \\
		$A_{\rm in}$ & Flux contrast of a feature with the regolith \\
		$\phi_0$, $\delta\phi$ & Mean incident solar flux and its perturbation [W m$^{-2}$] \\
		$C_n$, $\omega$ & Harmonic coefficients of $\delta\phi$ [W m$^{-2}$]; fundamental angular frequency [rad s$^{-1}$] \\
		$S_\phi$, $A_\phi$ & PSD and ASD of the solar-flux perturbation \\
		$k$ & Radial (Hankel) wavenumber [m$^{-1}$] \\
		$\alpha$, $\alpha_n$ & Complex vertical decay constant (for harmonic $n$) [m$^{-1}$] \\
		$A_0(k)$ & Hankel coefficients of the surface forcing \\
		$J_0$ & Bessel function of the first kind, order 0 \\
		$\alpha_T$ & Coefficient of linear thermal expansion [K$^{-1}$] \\
		$\nu$ & Poisson ratio of the regolith \\
		$u_z$ & Vertical surface displacement [m] \\
		$\theta_x,\theta_y$ & Tilt components of the seismometer [rad] \\
		$g_{\rm Moon}$ & Lunar surface gravity [m s$^{-2}$] \\
		$h_r$ & Linearized radiative coefficient, $4\sigma T_0^3$ [W m$^{-2}$ K$^{-1}$] \\
		$\varepsilon_M$, $\varepsilon_g$ & Infrared emissivity of the MLI shield and of the regolith \\
		$\eta$ & Fraction of the solar-flux fluctuation transmitted by the MLI shield \\
		$R_b$, $h_b$ & Radius and height of the instrument body [m] \\
		$\zeta$, $\Delta$ & Solar zenith angle [rad]; shift of the body shadow, $h_b\tan\zeta$ [m] \\
	\end{tabular}
	\section{Introduction}
	
	The Apollo geophysical experiments remain the primary foundation for our present understanding of the Moon’s interior \cite{latham1970,langseth1976}. Measurements from the Apollo seismic network, magnetometers, and heat-flow probes showed that the lunar subsurface preserves information about the Moon’s seismic activity, thermal evolution, and internal structure \cite{nakamura1982,langseth1976,heiken1991,khan2002,lognonne+2003}. See \cite{lognonne2015,Garcia2019} for reviews.  The deployment of these geophysical payloads on the lunar surface also demonstrated that successful collection of planetary geophysical data depends not only on instrument sensitivity, but also on a careful understanding of the surface environment and mitigation strategy of the sources of noises related to such surface deployment, as illustrated also on Mars with the InSight mission and its seismometer\cite{Banerdt2020,lognonne2019}. 
    Both the Apollo and InSight seismometers were therefore shielded. These shields were designed in order to generate a "portable" seismic vault, protecting the seismometers from thermal and, on Mars, wind effects. See for Mars  \cite{mimoun2017,Pinot2024}. Despite these efforts, the horizontal components of seismometers on Mars \cite{Lognonn2020,Lognonn2023,Pinot2024} and the Moon \cite{lognonne2015} remain systematically more noisy as the vertical one. Such larger horizontal noise is generally mostly due to tilt noise, associated on Earth \cite{Beauduin1996,Zurn2021} and Mars \cite{Garcia2020} to surface pressure fluctuations, but also to surface temperature change, especially on airless terrestrial bodies like the Moon. 
    This paper focus on tilt noise associated to surface temperature, and is a first step toward its complete modeling. Tilt noise from surface temperature change has been nicely illustrated on Mars by the recording by SEIS of signals associated to the partial eclipse of the Sun by Phobos \cite{stahler2020}. SEIS detected here tilts of about 50 $nm/s^2$ during 30 sec Phobos eclipses generating 3K ground surface variations \cite{stahler2020}, which translates to 16 $pm/s^2$ per mK of surface variations and appears therefore as a significant source of noise for pico-g or better sensitivities.\\
    We will focus in this paper on the Moon, for which extreme sensitivities will be requested, especially for gravitational wave detections \cite{Harms_2021,branchesi2023,jani2025laserinterferometerlunarantenna} and normal modes observations \cite{Panning2026}. Furthermore its surface is exposed directly to solar radiation and space without atmospheric buffering and as this will be therefore one of the primary source of stochastic thermo-elastic noise. Over the lunar day, the upper regolith undergoes large temperature changes, while the low thermal conductivity of the soil confines much of the diurnal thermal response to shallow depths \cite{carrier1991,paige2010,vasavada2012,hayne2017}. These cyclic temperature variations can produce expansion and contraction of the near-surface material, generating thermoelastic deformation, tilt, and thermal disturbances that may be recorded by surface instruments. Therefore, future lunar seismic and gravitational waves sensors must be interpreted in the context of both the natural thermal cycle and local modifications produced by nearby surface features. For the proposed Lunar Gravitational-wave Antenna, for example, it has been recognized that stations outside permanently shadowed regions would need to be buried to avoid excess noise from ground tilt driven by temperature variations \cite{ajith2025lgwa}. But how much can be achieved for surface installation protected by thermal shield? And even in PSR, how much thermoelastic noise can be generated from sun indirect illumination \cite{Yin2024} ? 
	
	Local surface heterogeneities, such as rocks, lander components, shields, and nearby instruments, can perturb the natural diurnal heat flux and generate environmentally driven disturbances that complicate the interpretation of thermal measurements \cite{hayne2017,grott2010,nagihara2018}. They modify the surface energy balance by altering local albedo, emissivity, and thermal inertia. These changes perturb both the amplitude and phase of the diurnal temperature cycle of the underlying regolith \cite{hayne2017}. Because the forcing is spatially nonuniform, it establishes lateral temperature gradients in addition to the vertical heat conduction traditionally assumed in one-dimensional models. In addition of bias on the apparent heat flux measured by thermal probes \cite{grott2010}, such gradients can impart thermoelastic stresses within the near-surface regolith, leading to differential ground deformation affecting seismometers tilts. Unlike transient shadowing, permanent shadows created by spacecraft structures, such as thermal blankets or regions beneath landers, can impose persistent modifications to the subsurface thermal regime, with cumulative effects extending over multiple diurnal cycles \cite{nagihara2018}. To mitigate these influences, instruments are typically positioned at prescribed distances from landers and shielded regions to reduce conductive and radiative interference, ensuring that measurements more accurately reflect the natural heat flow from the lunar interior rather than spacecraft-induced artifacts \cite{langseth1976}. 
	
	In this study, we adopt a perturbation-based formulation that isolates the lateral thermal response induced by localized surface heterogeneities. Rather than modeling the absolute lunar temperature field, we decompose the temperature as $T(r,z,t) = T_0(z,t) + \vartheta(r,z,t),$ where $T_0$ represents the laterally uniform background thermal state driven by diurnal solar forcing, and $\vartheta$ denotes the spatially localized temperature perturbation arising from solar flux time variability. By explicitly subtracting the homogeneous component of the solar forcing, the analysis focuses exclusively on the differential temperature field responsible for lateral heat diffusion, thermoelastic strain, and potential instrument bias. This approach enables a geometry-driven assessment of thermal shielding effects that is, to first order, decoupled of absolute surface temperatures computations. We first derive a closed-form analytical solution for the subsurface temperature perturbation beneath a single circular insulating footprint, which serves as a canonical representation of an isolated surface heterogeneity such as a seismometer thermal shield or other nearby features. We then extend the formulation to multiple insulating regions distributed across the surface, allowing the combined influence of nearby features and shielding elements to be evaluated through linear superposition. The resulting framework provides a physically transparent and computationally efficient means of quantifying how localized surface features modify the shallow thermal field and induce thermoelastic deformation relevant to the stability and interpretation of lunar seismic and thermal measurements. The Supplementary Material gives the derivations, the numerical inputs and additional figures on the shadow of the instrument body (Supplementary Section~\ref{app:noon_shadow}).
	
	\section{Physical Setting and Modeling Assumptions}
	
	\begin{figure}[h!]
		\centering
		\includegraphics[width = 1\textwidth,keepaspectratio]{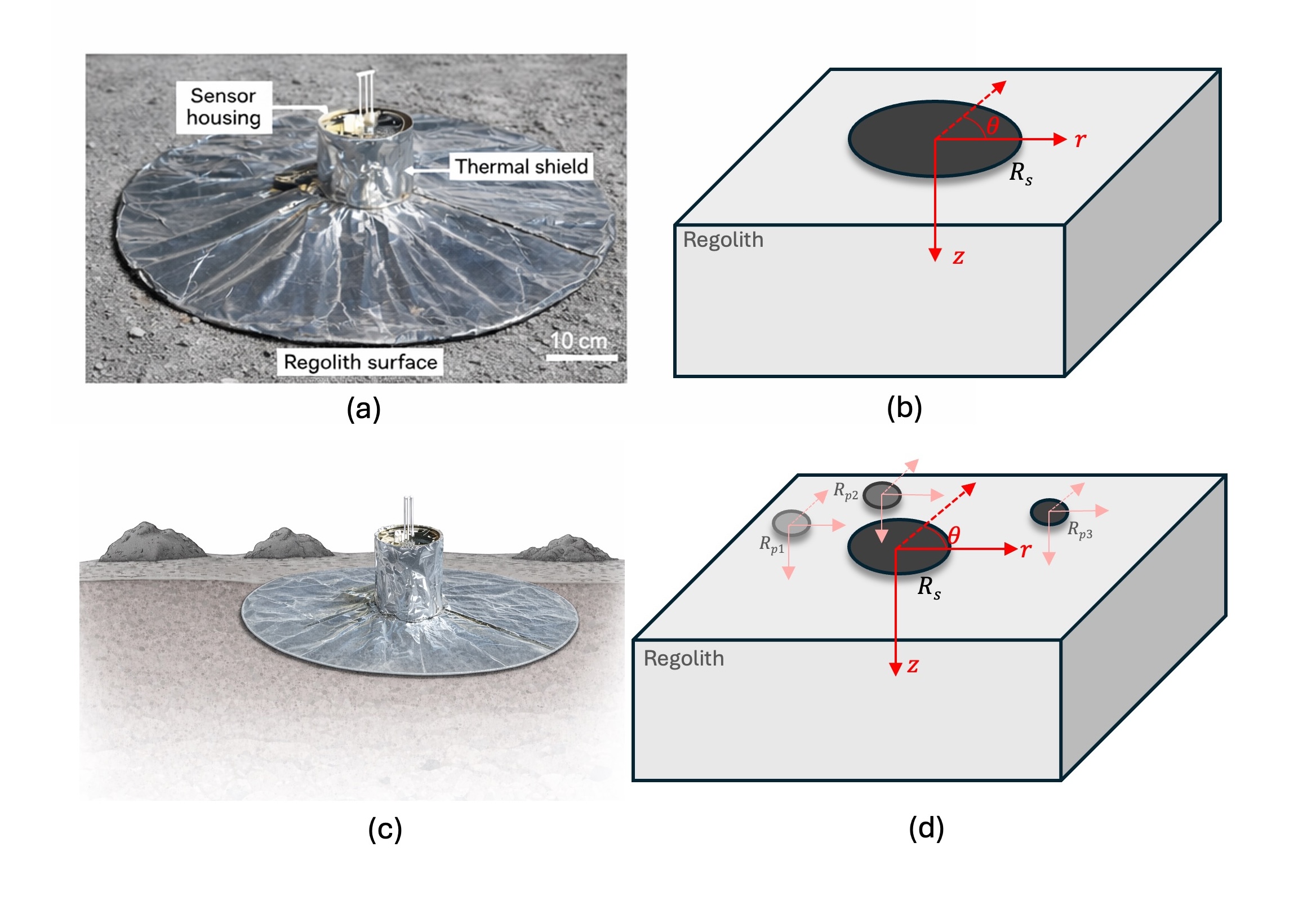}
		\captionsetup{font={small,stretch=0.8}}
		\caption{ (a) Apollo lunar seismometer installation showing the central sensor housing and surrounding Mylar thermal shield on the regolith surface. (b) Idealized axisymmetric model used in this study, with a circular thermal shield of radius $R_s$ resting on a semi-infinite regolith half-space. (c) Conceptual seismometer-shield configuration with nearby surface heterogeneities. (d) Extension to multiple localized surface features of radii $R_{pi}$. }
		\label{fig_geometry}
	\end{figure}
	
	On planetary missions, scientific instruments and lander components are commonly covered with
	protective Mylar thermal blankets \cite{NASA_LunarLander_ThermalProtection, lognonne2019} to shield them from extreme surface temperature variations (Figure \ref{fig_geometry}a).
	These thermal shields regulate heat exchange with the environment primarily by limiting
	radiative transfer at the surface and thereby reducing the penetration of temperature
	variations into the subsurface regolith.
	\par The problem is formulated in an axisymmetric cylindrical coordinate system
	$(r,\theta,z)$, with the origin located at the center of the thermal shield on the regolith
	surface (Figure~\ref{fig_geometry}b). The $z$-axis is defined as positive downward into the
	regolith. In the absence of localized surface heterogeneities, all fields are assumed to be
	independent of the azimuthal coordinate $\theta$, reducing the problem to two dimensions
	$(r,z)$.
	
	The regolith is modeled as a semi-infinite half-space occupying $z \geq 0$ and is assumed to
	behave as a homogeneous, isotropic, linear thermoelastic solid. Material properties including
	density, thermal conductivity, heat capacity, elastic moduli, and thermal expansion
	coefficient are taken as constant with depth. This idealization isolates the fundamental
	physics of thermoelastic coupling while remaining appropriate for the shallow depths that
	control seismometer tilt and near-surface deformation.
	
	In addition to the thermal shield, discrete surface features such as exposed rocks, or other instrument components act as
	localized thermal anomalies (Figure~\ref{fig_geometry}c). These surface features absorb and reradiate solar energy
	differently from the surrounding regolith, producing spatially heterogeneous surface heat
	fluxes. Each feature is idealized as a circular region of radius $R_{p_i}$ located at a surface
	position $(r_{p_i}, \theta_{p_i})$ relative to the seismometer (Figure~\ref{fig_geometry}d).
	The thermal forcing associated with these features introduces azimuthal variability in the
	surface boundary condition, locally breaking axisymmetry. Despite this local three-dimensionality, the thermoelastic response induced by each feature is
	computed independently and superposed, owing to the linearity of the governing heat
	conduction and elasticity equations. This approach allows the combined influence of multiple
	features to be evaluated while retaining the computational
	efficiency and physical transparency of the axisymmetric formulation. 
	\subsection{Thermal Shield Geometry and Boundary Conditions}
\begin{figure}[h!]
		\centering
		\includegraphics[width = 1\textwidth,keepaspectratio]{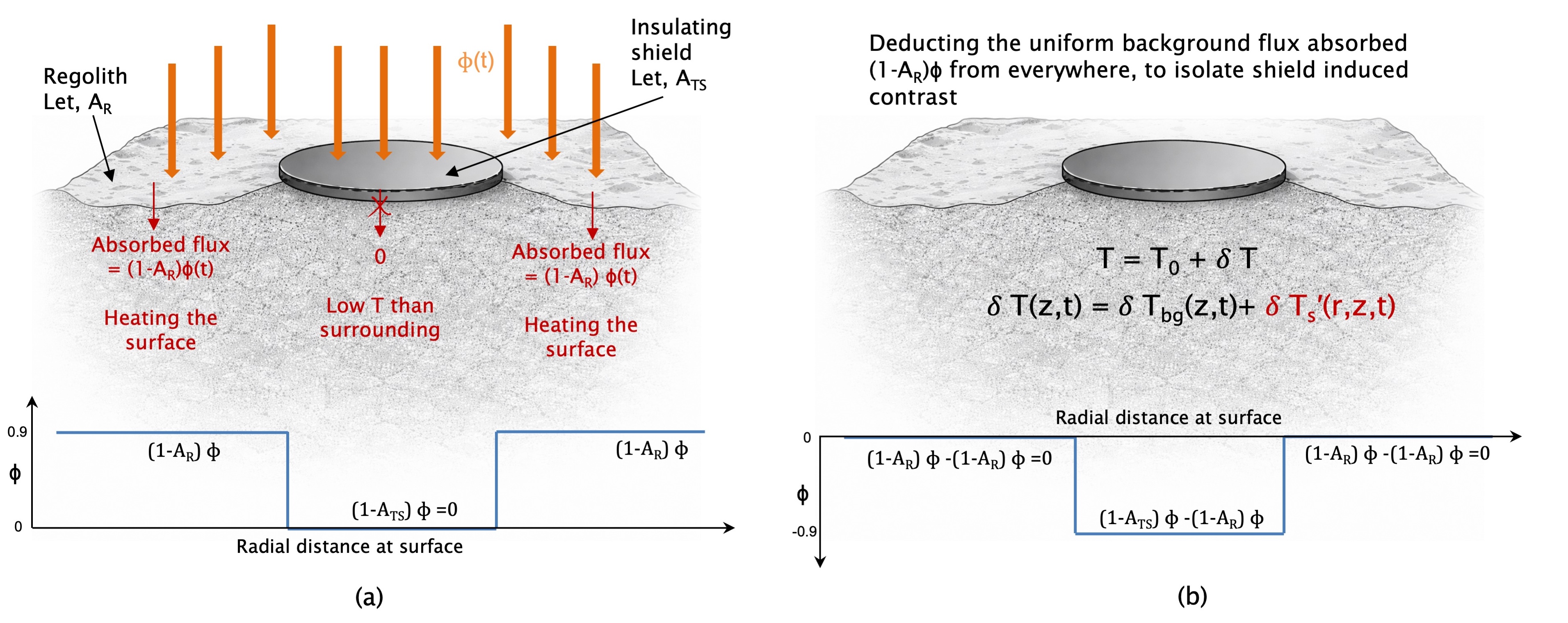}
		\captionsetup{font={small,stretch=0.8}}
		\caption{
			Schematic of surface thermal forcing beneath a lunar thermal shield. (a) Differential solar heating between exposed and shielded regolith. (b) Subtraction of the uniform background forcing to isolate the shield-induced thermal perturbation responsible for lateral temperature gradients and thermoelastic deformation.
		}
		\label{fig:flux}
	\end{figure}

	The thermal shield is idealized as a circular disk of radius $R_s$ placed directly on the
	regolith surface (Figure~\ref{fig:flux}a). The shield is assumed to block the incident sunlight, so that the regolith surface
	beneath it absorbs negligible solar flux. This is modeled as an effective flux mask over the
	shielded region, $r \leq R_s$: the absorbed solar flux is removed there, while the same
	linearized radiative-conductive surface energy balance is retained at every $r$. 
	\par Outside the shielded region ($r > R_s$), the regolith surface is directly exposed to incident
	solar radiation. The net absorbed surface heat flux is given by $(1-A_R)\,\phi(t)$
	where $A_R$ is the regolith albedo and $\phi(t)$ denotes the time-dependent incident solar
	flux. As illustrated in Figure~\ref{fig:flux}a, this produces surface heating outside the
	shield and a relative cooling beneath it, generating strong lateral temperature gradients at
	the shield edge.
	
	To isolate the thermoelastic response driven specifically by the presence of the thermal
	shield, the spatially uniform background heating that would occur in the absence of the
	shield is subtracted. This uniform component does not generate lateral temperature gradients
	and therefore does not contribute to differential deformation or tilt. After subtraction,
	the surface forcing represents only the shield-induced flux contrast
	(Figure~\ref{fig:flux}b).
	
	The resulting surface heat-flux perturbation is expressed as
	\begin{equation}
		q_s(r,t) =
		\begin{cases}
			(1-A_{TS})\,\phi(t) - (1-A_R)\,\phi(t), & r \leq R_s, \\
			0, & r > R_s,
		\end{cases}
	\end{equation}
	where $A_{TS}$ is the effective albedo of the thermal shield. For an ideal insulating shield,
	$(1-A_{TS}) \approx 0$, reducing the forcing beneath the shield to a negative flux anomaly
	equal to $-(1-A_R)\phi(t)$.
	
	With this formulation, the total temperature field is decomposed as
	\[
	T(r,z,t) = T_0(z) + \delta T_{bg}(z,t) + \delta T_s(r,z,t),
	\]
	where $T_0(z)$ is the steady background temperature profile, $\delta T_{bg}$ represents the
	one-dimensional temperature response to uniform solar forcing, and $\delta T_s$ is the
	shield-induced perturbation that varies radially and with depth. Only $\delta T_s$ contributes
	to lateral AC temperature gradients and drives the thermoelastic deformation and seismometer tilt
	analyzed in this study.

	\subsection{Imperfect multilayer-insulation shield}
	\label{sec:mli_model}
	Real thermal shields are not perfectly opaque. The Apollo shield is a multilayer insulation (MLI) blanket, whose infrared emissivity is $\varepsilon_M\approx0.02$. When sunlight warms the top of the blanket, part of that heat is radiated from its underside to the regolith. For two large parallel surfaces of emissivities $\varepsilon_g$ (regolith) and $\varepsilon_M$ (MLI), the net radiative exchange is \cite{bergman2011}
	\begin{equation}
		q=\eta\,\sigma\left(T_1^4-T_2^4\right),\qquad
		\eta=\frac{1}{1/\varepsilon_g+1/\varepsilon_M-1},
		\label{eq:eta}
	\end{equation}
	so that the ground beneath the shield receives a fraction $\eta$ of the fluctuation of the absorbed solar flux. With $\varepsilon_g=0.95$, $\eta=1/50$ for $\varepsilon_M=0.02$ and $\eta=1/10$ for a dust-degraded blanket with $\varepsilon_M=0.1$. The same factor reduces the radiative loss of the ground beneath the shield, which radiates to the underside of the blanket rather than to space, so that the linearized boundary condition there becomes
	\begin{equation}
		\eta\,(1-A_R)\,\delta\phi=\eta\,h_r\,\vartheta-\kappa\frac{\partial\vartheta}{\partial z},\qquad h_r=4\sigma T_0^3 .
		\label{eq:bc_mli}
	\end{equation}
	The weaker radiative coupling ($\eta h_r$ instead of $h_r$) increases the temperature response beneath the shield, mainly at low frequency where radiation dominates conduction: in one dimension the amplification is a factor $\approx5$ at 0.01~mHz, 2.2 at 0.1~mHz and 1.35 at 1~mHz.

	The instrument itself stands on the shield and is opaque. Its body, of radius $R_b$ and height $h_b$, shades the blanket beneath it, so that no flux leaks through that part. With a leaky shield the ground therefore has three zones: outside the shield it receives the full fluctuation, beneath the sunlit part of the shield the fraction $\eta$, and beneath the body shadow nothing, when the fluctuations of the operating power are small enough to be neglected. Writing $m_s$ and $m_b$ for the indicator functions of the shield and of the body shadow (1 inside, 0 outside), the fraction of the fluctuation reaching the ground is
	\begin{equation}
		f=(1-m_s)+\eta\,(m_s-m_b)=(1-\eta)(1-m_s)+\eta\,(1-m_b).
		\label{eq:three_zones}
	\end{equation}
	The leaky shield is therefore equivalent to an opaque shield weakened by $(1-\eta)$, plus a small copy, of weight $\eta$, of the body shadow alone without any shield. Both are circular features to which Eq.~(\ref{eq:uz_feature}) applies, the second being computed with the radiative coefficient $\eta h_r$ of Eq.~(\ref{eq:bc_mli}).

	Let us consider an ideal case where we neglect any tilt due to the instrument itself and where the instrument has been leveled. When the Sun is at the zenith, the body shadow lies directly beneath the instrument. Because the instrument stands on its own tripod, the shadow is then centered on the legs and the temperature induced displacement are the same at the three legs. No tilt is then generated unless the symmetry is broken, for example by differences in the ground thermal conductivity below the legs or albedo. When the Sun is at a zenith angle $\zeta$, the shadow is displaced away from the Sun by
	\begin{equation}
		\Delta\approx h_b\tan\zeta,
		\label{eq:shadow_shift}
	\end{equation}
	the legs are no longer equidistant from its edge, and the body term produces a tilt. In the axisymmetric framework the displaced shadow is represented by the disk of radius $R_b$ moved by $\Delta$. In reality the ground directly beneath the body remains dark and the shadow is elongated on the side away from the Sun, so only one side of the dark region moves; the shifted disk is therefore expected to overestimate this tilt.

	\section{Governing Equations}
	
	We model heat transport in the regolith using the heat conduction equation
	\begin{equation}
		\rho C_p \, \frac{\partial T}{\partial t}
		=
		\kappa \nabla^2 T,
	\end{equation}
	where $\rho$ is the regolith density, $C_p$ is the specific heat capacity, and $\kappa$ is the
	thermal conductivity. Heat transfer is assumed to occur solely by conduction within the
	regolith, consistent with the absence of an atmosphere on the Moon.
	
	At the regolith surface ($z=0$), the temperature evolution is governed by an energy balance
	between absorbed solar radiation, thermal radiation to space, and conductive heat flux into
	the subsurface. This balance may be written as
	\begin{equation}
		(1-A)\,\phi(t)
		=
		\sigma \left(T^4 - T_{sky}^4\right)
		-
		\kappa \frac{\partial T}{\partial z}\bigg|_{z=0},
	\end{equation}
	where $\phi(t)$ is the incident solar flux, $A$ is the surface albedo, $\sigma$ is the
	Stefan-Boltzmann constant, $T$ is the surface temperature, and $T_{sky}$ is the effective sky
	temperature. The final term represents conductive heat loss into the regolith.
	
	\subsection{Linearization About the Equilibrium State}
	
	We decompose the temperature and solar forcing into steady and time-varying components,
	\begin{equation}
		T = T_0 + \vartheta, \qquad
		\phi = \phi_0 + \delta \phi,
	\end{equation}
	where $T_0$ and $\phi_0$ denote the equilibrium surface temperature and mean solar flux,
	respectively, and $\vartheta$ and $\delta \phi$ represent small perturbations about this
	equilibrium.
	
	Substitution into the heat equation yields the linearized governing equation for the
	temperature perturbation,
	\begin{equation}
		\rho C_p \frac{\partial \vartheta}{\partial t}
		=
		\kappa \nabla^2 \vartheta.
		\label{eq:heat_lin}
	\end{equation}
	
	At the surface, substitution into the energy balance equation gives
	\begin{equation}
		(1-A)(\phi_0 + \delta \phi)
		=
		\sigma \left[(T_0 + \vartheta)^4 - T_{sky}^4\right]
		-
		\kappa \frac{\partial (T_0 + \vartheta)}{\partial z}\bigg|_{z=0}.
	\end{equation}
	Expanding the radiative term to first order, and neglecting higher-order terms yields
	\begin{equation}
		(1-A)\phi_0 + (1-A)\delta \phi
		=
		\sigma \left(T_0^4 - T_{sky}^4\right)
		+
		4\sigma T_0^3 \vartheta
		-
		\kappa \frac{\partial T_0}{\partial z}
		-
		\kappa \frac{\partial \vartheta}{\partial z}.
	\end{equation}
	
	The equilibrium temperature $T_0$ satisfies the steady-state balance
	\begin{equation}
		(1-A)\phi_0
		=
		\sigma \left(T_0^4 - T_{sky}^4\right),
	\end{equation}
	and, for a horizontally uniform equilibrium state, the background temperature gradient
	vanishes at the surface, $	\frac{\partial T_0}{\partial z}\bigg|_{z=0} = 0.$
	Subtracting the equilibrium balance, the linearized surface boundary condition becomes
	\begin{equation}
		(1-A)\,\delta \phi
		=
        \sigma \left(T_0^4 - T_{sky}^4\right) {\delta \phi \over \phi_0}
        =
		4\sigma T_0^3 \vartheta
		-
		\kappa \frac{\partial \vartheta}{\partial z}.
		\label{eq:bc_lin}
	\end{equation}
	
	This expression relates time-dependent variations in incident solar flux to the resulting surface temperature perturbation and conductive heat flux into the regolith. It provides the boundary condition used to solve for the harmonic thermal response and resulting thermoelastic deformation. Note also that this equation can be extended to Permanently Shadowed Regions (PSR) for which we can rewrite the equation while assuming that the effective albedo integrates the indirect lighting: 
    
    \begin{equation}
         { \left(T_0^4 - T_{sky}^4\right) \over T_0^4} {\delta \phi \over \phi_0}
        =
		4 {\vartheta \over T_0}
		-
		{ \kappa \over \sigma T_0^3} \frac{\partial }{\partial z} \left( {\vartheta \over T_0} \right).
	\end{equation}

    This suggests that for most of the PSRs, where the 40K temperature \cite{paige2010} remains much greater than the night sky temperature, the surface temperature fluctuations will only decrease linearly with the PSR average temperature but remains, with only about one order of magnitude less compared to Moon equator at noon.
	
	\section{Analytical Solution for a Localized Surface Perturbation}
	
	We seek an analytical solution for the temperature perturbation $\vartheta$ induced by a
	spatially localized, time-dependent surface heat-flux anomaly. The perturbation is assumed to
	depend on radial distance $r$, depth $z$, and time $t$, and is expressed in separable form as
	\begin{equation}
		\vartheta(r,z,t) = R(r)\,Z(z)\,e^{i\omega t},
	\end{equation}
	where $\omega$ is the angular frequency of the harmonic forcing. \\
	Substitution into the linearized heat equation~(\ref{eq:heat_lin}), and adopting cylindrical coordinates yields
	\begin{equation}
		\frac{1}{r}\frac{\partial}{\partial r}
		\left( r \frac{\partial \vartheta}{\partial r} \right)
		+
		\frac{\partial^2 \vartheta}{\partial z^2}
		-
		\frac{\rho C_p}{\kappa} i\omega \vartheta
		= 0,
	\end{equation}
	where axisymmetry has been assumed and all azimuthal derivatives vanish.
	
	Dividing through by $R(r)Z(z)$ separates the equation into radial and vertical components,
	\begin{equation}
		\frac{1}{R}\frac{1}{r}\frac{\mathrm{d}}{\mathrm{d}r}
		\left( r \frac{\mathrm{d}R}{\mathrm{d}r} \right)
		+
		\frac{1}{Z}\frac{\mathrm{d}^2 Z}{\mathrm{d}z^2}
		=
		\frac{\rho C_p}{\kappa} i\omega.
	\end{equation}
	
	Introducing a separation constant $\alpha^2$,
	\begin{equation}
		\frac{1}{Z}\frac{\mathrm{d}^2 Z}{\mathrm{d}z^2} = \alpha^2,
	\end{equation}
	the vertical solution that decays with depth is, $	Z(z) = e^{-\alpha z},$ where $\alpha$ is (a generally complex) vertical decay constant.\\
	Substitution into the radial equation yields
	\begin{equation}
		\frac{1}{r}\frac{\mathrm{d}}{\mathrm{d}r}
		\left( r \frac{\mathrm{d}R}{\mathrm{d}r} \right)
		+
		\left( \alpha^2 - \frac{\rho C_p}{\kappa} i\omega \right) R = 0.
	\end{equation}
	Defining
	\begin{equation}
		k^2 = \alpha^2 - \frac{\rho C_p}{\kappa} i\omega,
	\end{equation}
	this equation reduces to Bessel’s equation of order zero,
	\begin{equation}
		r^2 \frac{\mathrm{d}^2 R}{\mathrm{d}r^2}
		+
		r \frac{\mathrm{d}R}{\mathrm{d}r}
		+
		k^2 r^2 R = 0.
	\end{equation}
	
	The physically admissible solution finite at the origin is the Bessel function of the first
	kind,	$R(r) = J_0(kr).$ To accommodate arbitrary radial forcing, the solution is expressed using a zeroth-order
	Hankel transform,
	\begin{equation}
		R(r) = \int_0^\infty k\,R_0(k)\,J_0(kr)\,\mathrm{d}k,
	\end{equation}
	where $R_0(k)$ are the Hankel coefficients.
	
	Using the identity
	\begin{equation}
		\frac{1}{r}\frac{\mathrm{d}}{\mathrm{d}r}
		\left( r \frac{\mathrm{d}J_0(kr)}{\mathrm{d}r} \right)
		= -k^2 J_0(kr),
	\end{equation}
	each Hankel mode satisfies 	$k^2 = \alpha^2 - \frac{\rho C_p}{\kappa} i\omega,$
	implying
	\begin{equation}
		\alpha = \sqrt{k^2 + \frac{\rho C_p}{\kappa} i\omega}.
	\end{equation}
	Thus, the perturbation \( \vartheta(r,z,t) \) can be obtained in terms of radial functions as,
	\begin{equation}
		\vartheta(r,z,t) = \int_0^\infty k \, \, R_0(k) \, J_0(kr) e^{-\alpha z} e^{i\omega t} \mathrm{d}k,
		\label{eq:theta_hankel}
	\end{equation}
	where 	$\alpha(k)$ controls the penetration depth. A detailed derivation can be found in Supplementary Section~\ref{app:derivation}.
	
	\subsection{Surface Forcing and Boundary Condition}
	
	The spatial structure of the surface heat-flux perturbation is represented by a function
	$A(r)$, expressed via its Hankel transform,
	\begin{equation}
		A(r) = \int_0^\infty k\,A_0(k)\,J_0(kr)\,\mathrm{d}k.
	\end{equation}
	where $A_0$ is the Hankel coefficients of the sourcing term (here depends on the radius of the shield).
	At the surface, the linearized energy balance is given by Eq.~(\ref{eq:bc_lin}). Substituting $\vartheta(r,0)$ and $\partial \vartheta (r,0)/\partial z$ into Eq.~(\ref{eq:bc_lin}) yields,
	
	\begin{equation}
		\int_0^\infty k\, dk \left(\kappa \alpha + 4\sigma T_0^3 \right) R_0(k)\, J_0(kr)
		= \delta  \phi \int_0^\infty k\, dk \, A_0(k)\, J_0(kr)
	\end{equation}
	Because $\alpha=\alpha(k)$, the factor $\kappa\alpha+4\sigma T_0^3$ remains inside the integral. Equating the Hankel coefficients on both sides gives
	\begin{equation}
		R_0(k) = \frac{\delta  \phi \, A_0(k)}{\kappa \alpha + 4\sigma T_0^3}
	\end{equation}
	Substituting this back in Eq.~(\ref{eq:theta_hankel}) gives the temperature perturbation due to localized surface forcing term $\delta  \phi A(r)$,
	
	\begin{equation}
		\vartheta(r,z,t) =
		\delta \phi
		\int_0^\infty \frac{k \, A_0(k)}{\kappa \alpha + 4\sigma T_0^3} \, J_0(kr) \,
		e^{-\alpha z} e^{i\omega t} \, \mathrm{d}k
	\end{equation}
	
	For a general periodic forcing,
	\begin{equation}
		\delta \phi(t) = \sum_{n=1}^{\infty} C_n e^{i n \omega t},
	\end{equation}
	the full solution is obtained by linear superposition,
	\begin{equation}
		\vartheta(r,z,t)
		=
		\sum_{n=1}^{\infty}
		\int_0^\infty
		\frac{C_n\,A_0(k)}
		{\kappa \alpha_n + 4\sigma T_0^3}
		k\,J_0(kr)\,
		e^{-\alpha_n z}
		e^{i n \omega t}
		\,\mathrm{d}k,
	\end{equation}
	where
	\begin{equation}
		\alpha_n = \sqrt{k^2 + \frac{\rho C_p}{\kappa} i n \omega}.
	\end{equation}
The temperature field $\vartheta(r,z,t)$ is a departure from the laterally uniform diurnal thermal state caused only by the localized surface heterogeneities, not the absolute lunar temperature: it vanishes far from the perturbation source and should not be compared directly with Apollo heat-flow temperature records. 
	
	\subsection{Thermoelastic Coupling and Instrument Tilt}
		Under the assumption of linear, isotropic thermoelasticity, the thermal strain associated with the temperature perturbation is $\epsilon_{ij}=\alpha_T\,\vartheta\,\delta_{ij}$, where $\alpha_T$ is the coefficient of linear thermal expansion. The regolith is treated as a homogeneous elastic half-space with a traction-free surface. For each Hankel mode of the temperature field, $\vartheta\propto J_0(kr)\,e^{-\alpha z}$, the solution of the Navier equations gives a surface uplift $2(1+\nu)\,\alpha_T/(\alpha+k)$ per unit surface temperature (Supplementary Section~\ref{app:halfspace}). The vertical surface displacement produced by a feature is therefore
\begin{equation}
	u_z(r,t)
	=
	2(1+\nu)\,\alpha_T
	\sum_{n}
	C_n e^{in\omega t}
	\int_0^\infty
	\frac{k\,A_0(k)\,J_0(kr)}
	{\left(\kappa\alpha_n+4\sigma T_0^3\right)\left(\alpha_n+k\right)}
	\,\mathrm{d}k ,
	\label{eq:uz_feature}
\end{equation}
where $\nu$ is the Poisson ratio, and the displacement of the leg $\ell$ is $u_{z,\ell}=\sum_j u_{z,j}(d_{\ell j},t)$ summed over all features. For lateral scales much larger than the thermal skin depth ($k\ll|\alpha_n|$), Eq.~(\ref{eq:uz_feature}) reduces to a local relation, $u_z\simeq2(1+\nu)\alpha_T\int_0^\infty\vartheta\,\mathrm{d}z$, in which the displacement at a point depends only on the temperature directly below it.  The term in $k$ in Eq.~(\ref{eq:uz_feature}) denominator describes the elastic coupling of the regolith, the expansion of the sunlit ground outside the shield also displaces the ground beneath it, with an amplitude that decreases smoothly from the edge towards the centre, much more slowly than the temperature perturbation, which decays within a few skin depths of the edge. This coupling produces a gradient of $u_z$ across the instrument footprint and is the origin of the thermoelastic tilt.
	
	\section{Thermoelastic Bias in the Apollo 14 Seismometer from Surrounding Instrumentation}
	\begin{figure}[h!]
		\centering
		\includegraphics[width = 1\textwidth,keepaspectratio]{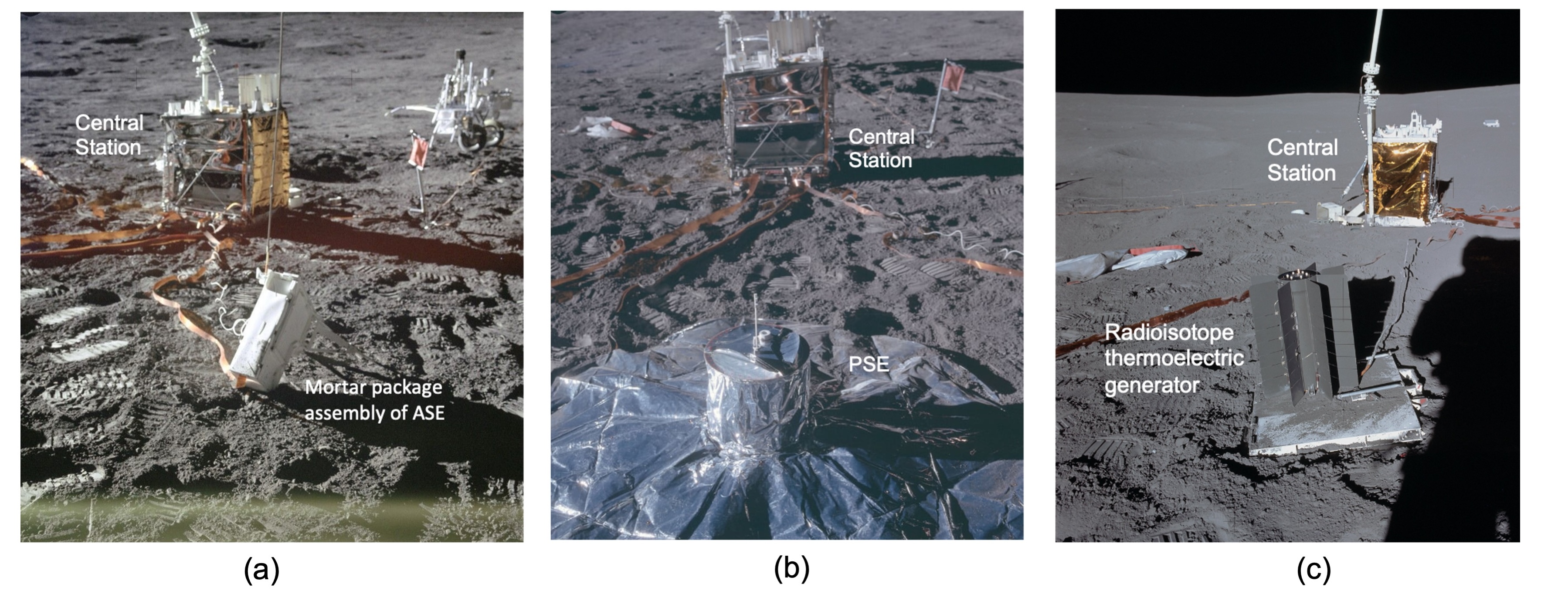}
		\captionsetup{font={small,stretch=0.8}}
		\caption{ Apollo 14 ALSEP deployment environment. The photographs (a), (b) and (c) show several thermally distinct components located near the Passive Seismic Experiment, including the Central Station, RTG, Mortar Package, and neighboring instrument packages. Image credit: NASA. }
		
		\label{fig_apollo1}
	\end{figure}
	We extend the analytical framework developed above to a mission-realistic configuration by incorporating the spatial arrangement of instruments deployed at the Apollo 14 landing site. Apollo Lunar Surface Experiments Package (ALSEP) represents a clustered environment (Figure \ref{fig_apollo1}) in which multiple thermally distinct components coexist within close proximity to the seismometer. Each of these components modifies the local surface energy balance through variations in albedo. The resulting thermal field is therefore governed by the superposition of multiple localized features, leading to lateral temperature gradients in the shallow regolith. These gradients, when coupled with thermoelastic deformation, can induce differential vertical motion across the seismometer legs and generate apparent tilt signals. By explicitly accounting for the geometry and thermal characteristics of surrounding ALSEP instruments, this section aims to quantify the extent to which such environmental interactions contribute to long-period tilt bias in the Passive Seismic Experiment.
	
	\subsection{Geometry of Apollo 14 ALSEP Layout}
	\label{sec:apollo_geometry}
	
\begin{figure}[h!]
		\centering
		\includegraphics[width = 1\textwidth,keepaspectratio]{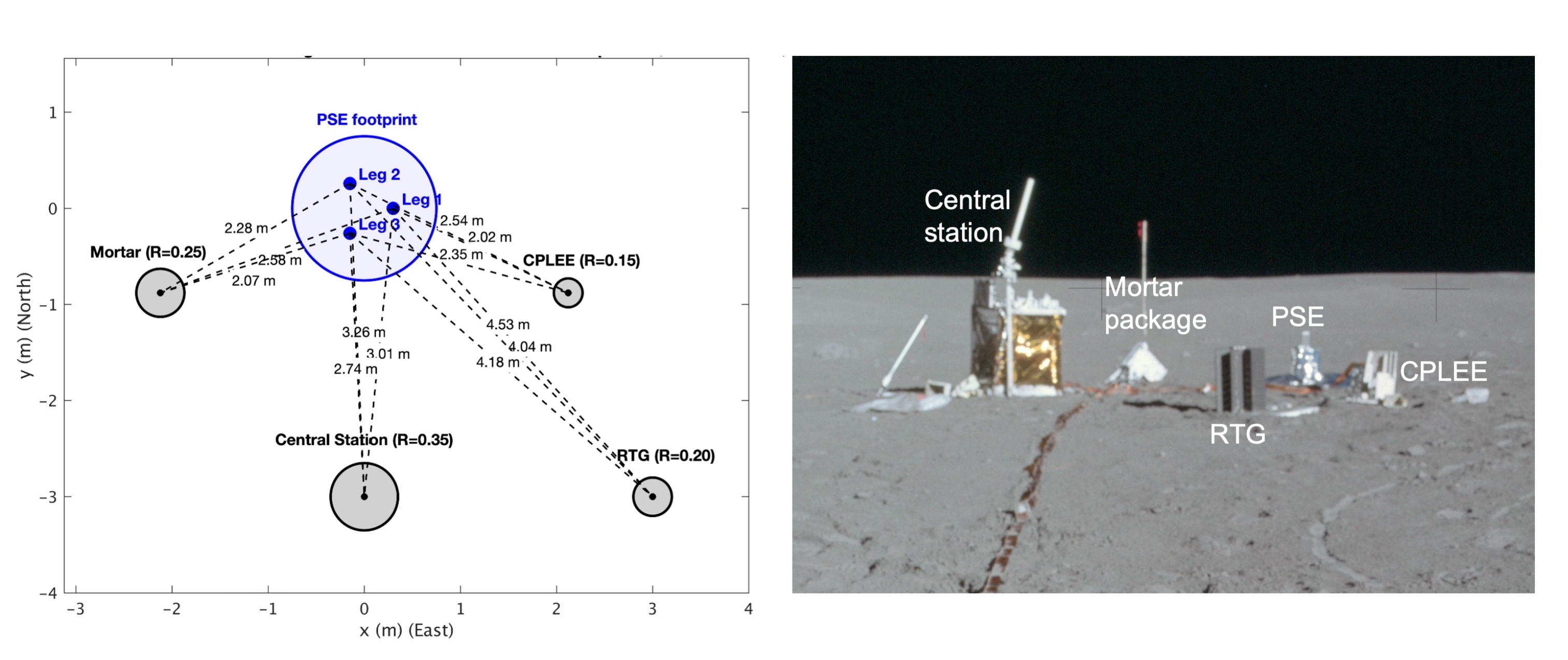}
		\captionsetup{font={small,stretch=0.8}}
		\caption{ Simplified Apollo 14 ALSEP geometry used in the thermoelastic tilt calculation. The PSE thermal shield is placed at the origin, and the three seismometer legs define the points where vertical displacement is sampled. Nearby ALSEP components are represented as localized circular surface features with effective radii and positions measured relative to the PSE. The right panel shows the corresponding Apollo deployment context. }
		\label{fig_apollo2}
	\end{figure}

	The simplified geometry of the ALSEP is shown in Figure \ref{fig_apollo2}.  The Passive Seismic Experiment (PSE) is taken as the reference location and is placed at the origin of a local Cartesian coordinate system on the regolith surface. The instrument is supported by a three-leg tripod, with the foot locations denoted by $\mathbf{x}_\ell = (x_\ell, y_\ell)$ for $\ell = 1,2,3$. The leg positions adopted are $\mathbf{x}_1=(0.120,0)$~m and $\mathbf{x}_{2,3}=(-0.080,\pm0.110)$~m. They form a nearly equilateral tripod, with sides of 0.22 and 0.23~m, whose centre lies 1.05~cm from the shield centre along $-x$, so that L1 is 0.120~m and L2, L3 are 0.136~m from the shield centre, and all three legs are 0.130~m from the tripod centre. We treat this as a centring error of $\approx1$~cm, representative of a manually deployed instrument; for an exactly centred tripod the contribution of the shield to the tilt cancels by symmetry (Section~\ref{sec:tilt_noise}). The instrument body is represented as an opaque disk of radius $R_b=0.14$~m and height $h_b=0.5$-0.6~m, centred on its own tripod (Section~\ref{sec:mli_model}); its shadow matters only when the shield transmits part of the flux (Section~\ref{sec:mli_apollo}). These points define the locations at which vertical surface displacement is sampled to compute instrument tilt. Lateral compliance and slip at the foot-regolith interface are neglected, such that the instrument
	tilt is governed entirely by differential vertical displacement of the three contact points.
	Surrounding the PSE are several ALSEP components, including the Central Station, the SNAP-27 radioisotope thermoelectric generator (RTG), the Charged Particle Lunar Environment Experiment (CPLEE), and the Mortar Package Assembly. Each component is represented as a localized surface feature centered at position $\mathbf{x}_j = (x_j, y_j)$, where $j = 1,2,\dots,N$.  The horizontal separation between the $\ell$-th seismometer leg and the $j$-th surface feature is defined as 	$d_{\ell j} = \|\mathbf{x}_\ell - \mathbf{x}_j\|$, which governs the contribution of each feature to the local thermoelastic response at the instrument.\\
	The reference calculations presented in the main text use the complete-masking limit: the regolith albedo is $A_R=0.12$ and every feature blocks the incident flux completely ($A_j=1$), so that the flux contrast is $A_{\rm in}=A_R-A_j=-0.88$ for every feature. This is the largest thermoelastic perturbation each instrument can produce. The PSE shield is first treated in the same way, as an opaque shield (Section~\ref{sec:tilt_noise}); the effect of its partial transparency is evaluated in Section~\ref{sec:mli_apollo}.

	\begin{figure*}[!t]
		\centering
		\begin{subfigure}[t]{0.47\textwidth}
			\centering
			\includegraphics[width=\linewidth]{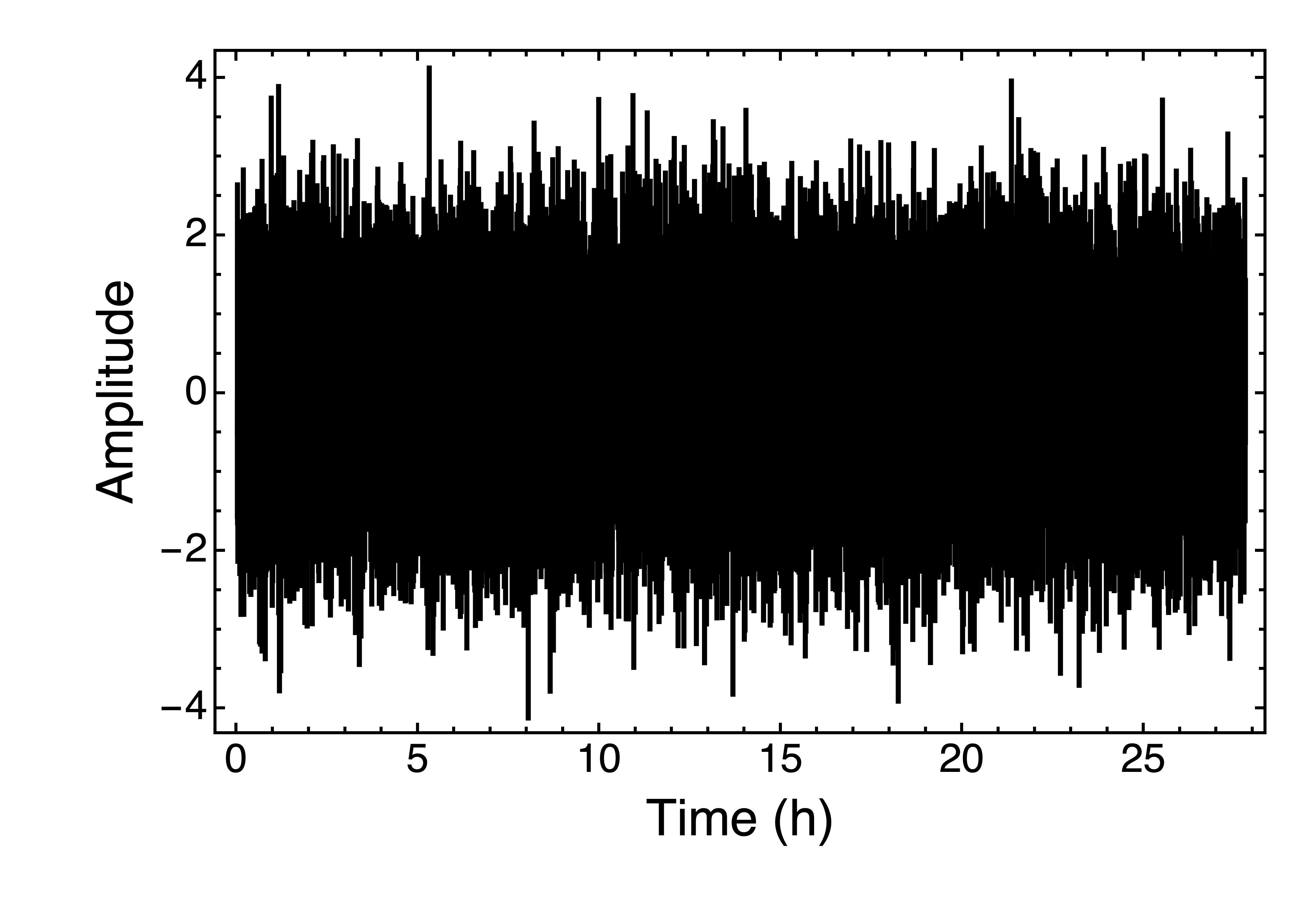}
			\caption{}
			\label{fig:p1}
		\end{subfigure}
		\hfill
		\begin{subfigure}[t]{0.47\textwidth}
			\centering
			\includegraphics[width=\linewidth]{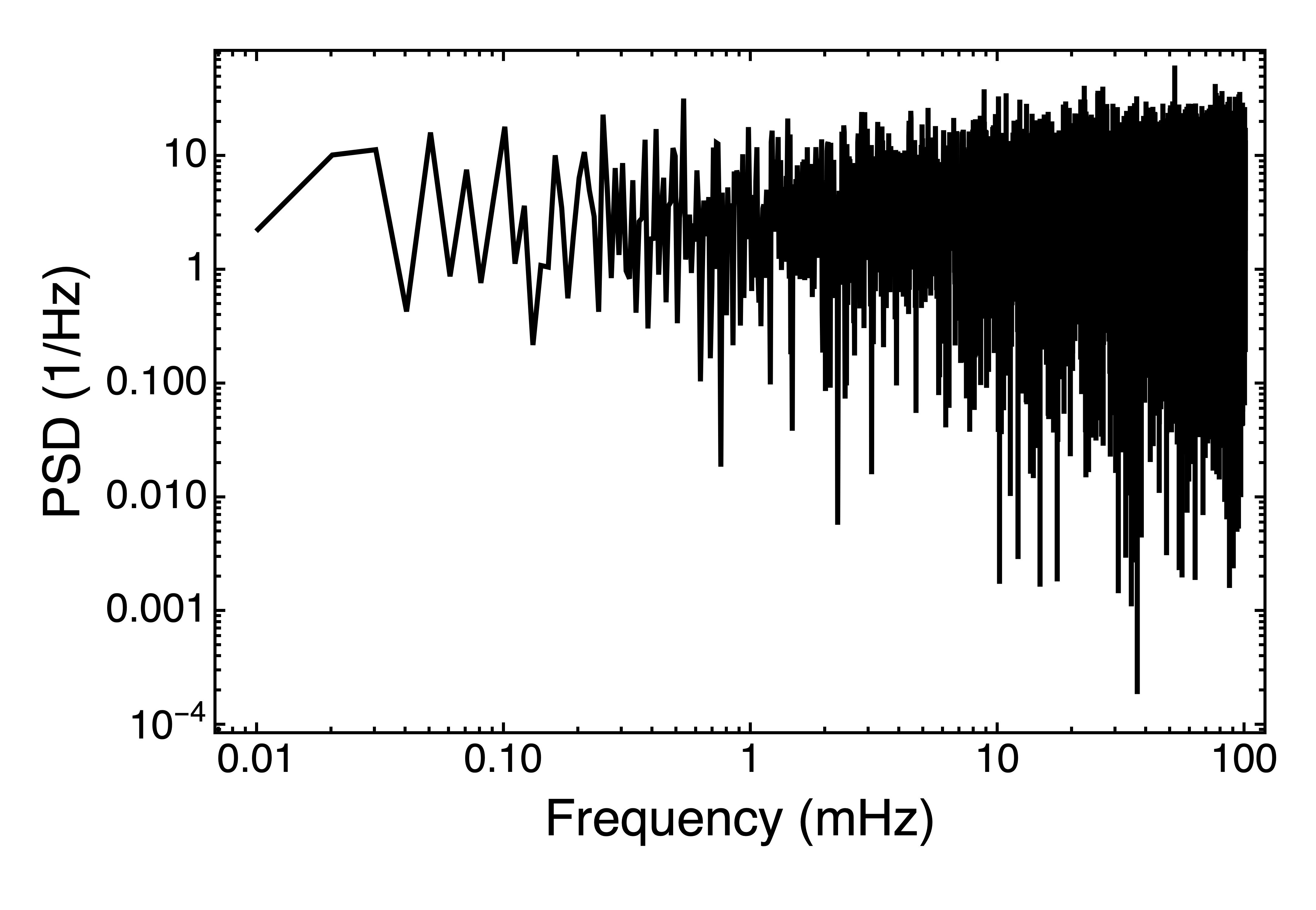}
			\caption{}
			\label{fig:p2}
		\end{subfigure}
		
		\vspace{0.01cm}
		
		\begin{subfigure}[t]{0.47\textwidth}
			\centering
			\includegraphics[width=\linewidth]{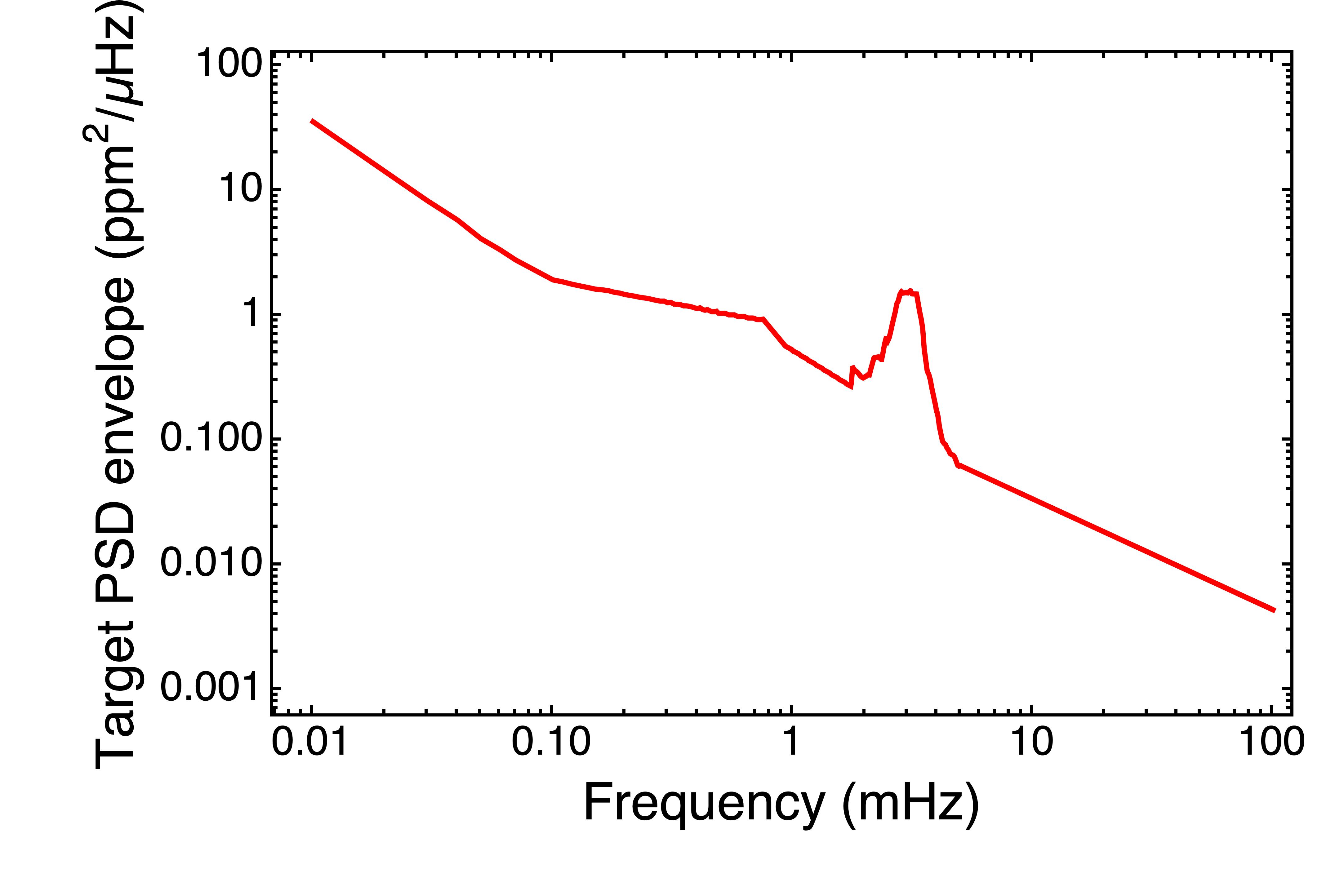}
			\caption{}
			\label{fig:p3}
		\end{subfigure}
		\hfill
		\begin{subfigure}[t]{0.47\textwidth}
			\centering
			\includegraphics[width=\linewidth]{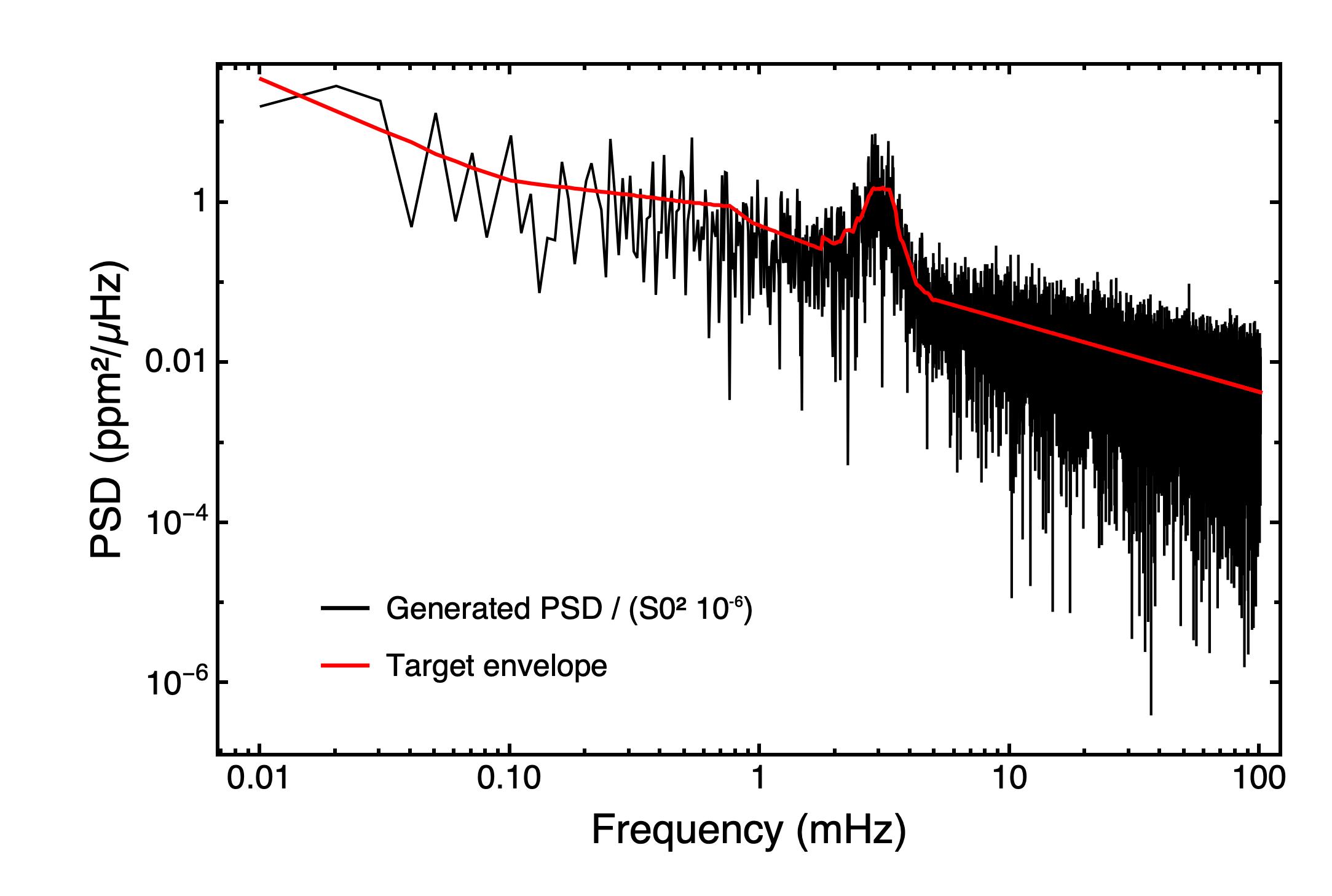}
			\caption{}
			\label{fig:p4}
		\end{subfigure}
		
		\vspace{0.01cm}
		
		\begin{subfigure}[t]{0.47\textwidth}
			\centering
			\includegraphics[width=\linewidth]{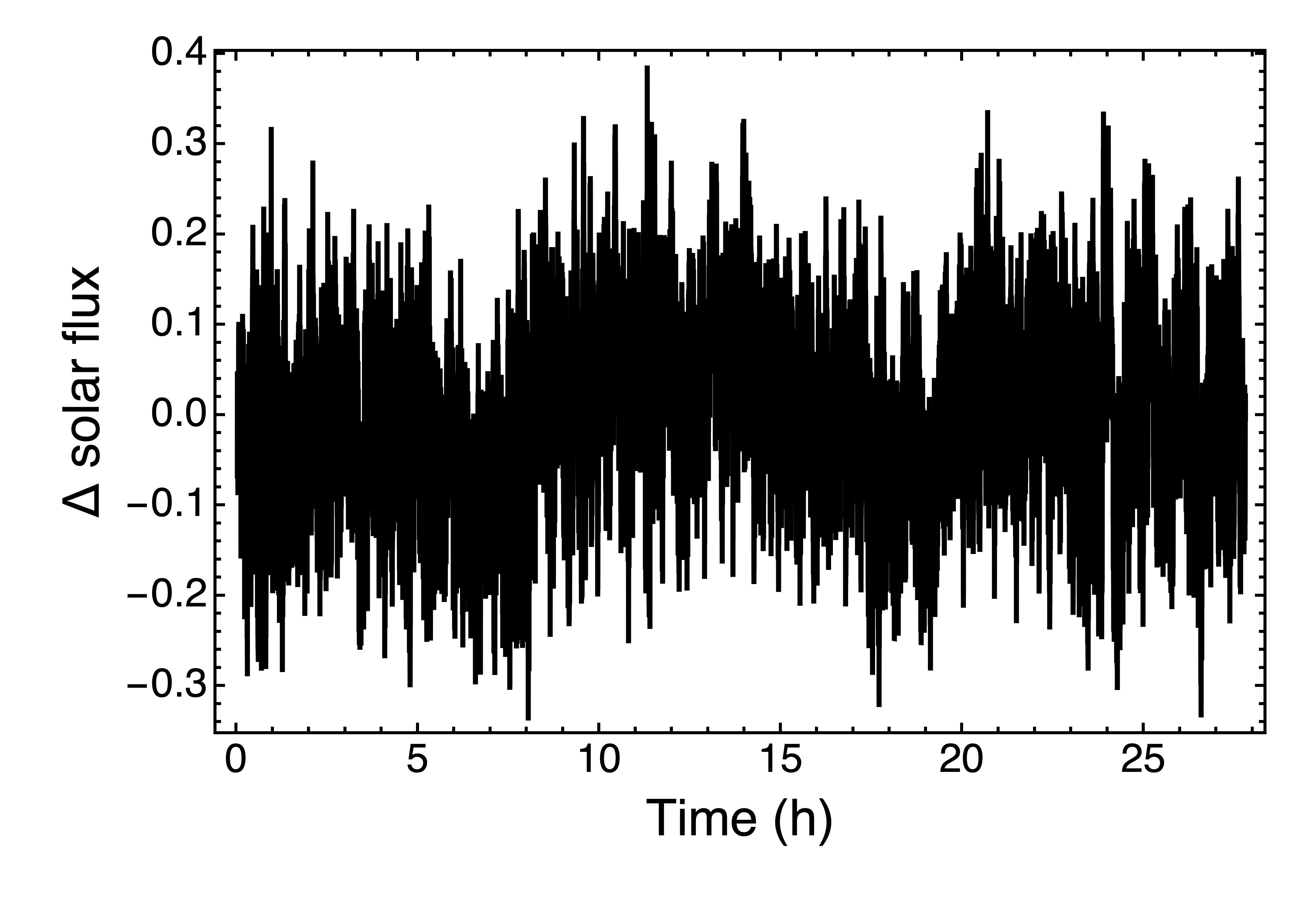}
			\caption{}
			\label{fig:p5}
		\end{subfigure}
		
		\caption{
			Generation of the random solar-flux perturbation.
			(a) Random seed time history.
			(b) Power spectral density of the random seed.
			(c) Target PSD envelope.
			(d) Comparison of the generated PSD with the target envelope, both expressed in $\mathrm{ppm^2/\mu Hz}$ (the generated PSD in $\mathrm{W^2\,m^{-4}\,Hz^{-1}}$ divided by $S_0^2\times10^{-6}$, $S_0=1360~\mathrm{W\,m^{-2}}$).
			(e) Time history of the resulting random solar-flux perturbation.
		}
		\label{fig:random_flux_generation}
	\end{figure*}
	
	\subsection{Solar-Flux Forcing}
	\label{sec:solar_forcing}
	
	The incident solar flux is decomposed into a mean component and a zero-mean
	perturbation,
	\begin{equation}
		\phi(t)=\phi_0+\delta\phi(t),
		\qquad
		\left\langle\delta\phi(t)\right\rangle=0,
	\end{equation}
	where $\phi_0$ determines the equilibrium surface temperature ($T_0=250$~K) and
	$\delta\phi(t)$ drives the time-dependent thermal response. The spectral
	shape of $\delta\phi(t)$ was obtained from the solar-irradiance variability
	reported by \cite{frohlich2004solar}. The digitized
	spectrum was interpolated and
	used as the target one-sided power spectral density, 	$S_{\phi}(f)	\quad [\mathrm{W^2\,m^{-4}\,Hz^{-1}}],
$	with the corresponding amplitude spectral density
	$A_{\phi}(f)=\sqrt{S_{\phi}(f)}$ in
	$\mathrm{W\,m^{-2}\,Hz^{-1/2}}$. 
	
	A stochastic solar-flux realization was generated in the frequency domain
	over the interval $		0.01~\mathrm{mHz}
		\leq f \leq
		100~\mathrm{mHz}.$
	Random phases were assigned to the positive-frequency coefficients, whose	magnitudes were scaled to reproduce $S_{\phi}(f)$. 
		Figure~\ref{fig:random_flux_generation} illustrates the construction of the
	stochastic solar flux realization. The initial random realization and its PSD are shown in
	Figs.~\ref{fig:p1} and \ref{fig:p2}, respectively. The target spectral envelope
	(Fig.~\ref{fig:p3}) is imposed on the random realization, resulting in a
	spectrum that follows the prescribed frequency dependence
	(Fig.~\ref{fig:p4}). The corresponding time-domain perturbation
	(Fig.~\ref{fig:p5}) is subsequently used as the solar flux variation as forcing in the model. Details of the digitized spectrum, Fourier normalization, stochastic reconstruction, and frequency resolution are provided in Supplementary Section~\ref{app:stochastic_forcing}.
			\begin{figure*}[!h]
	\centering
	\begin{subfigure}[h]{0.47\textwidth}
		\centering
		\includegraphics[width=\linewidth]{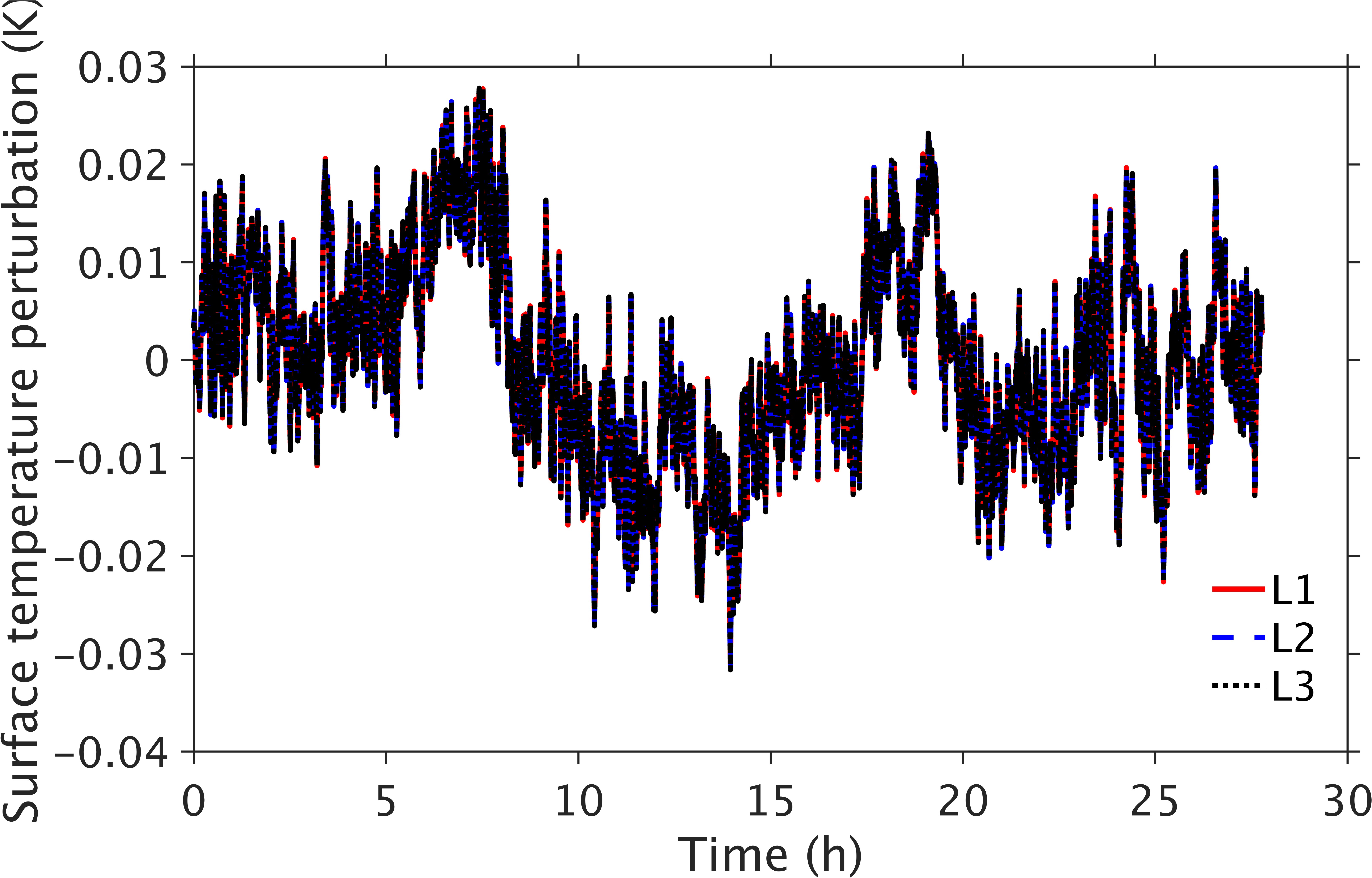}
		\caption{}
		\label{fig:temperature_legs}
	\end{subfigure}
	\hfill
	\begin{subfigure}[h]{0.47\textwidth}
		\centering
		\includegraphics[width=\linewidth]{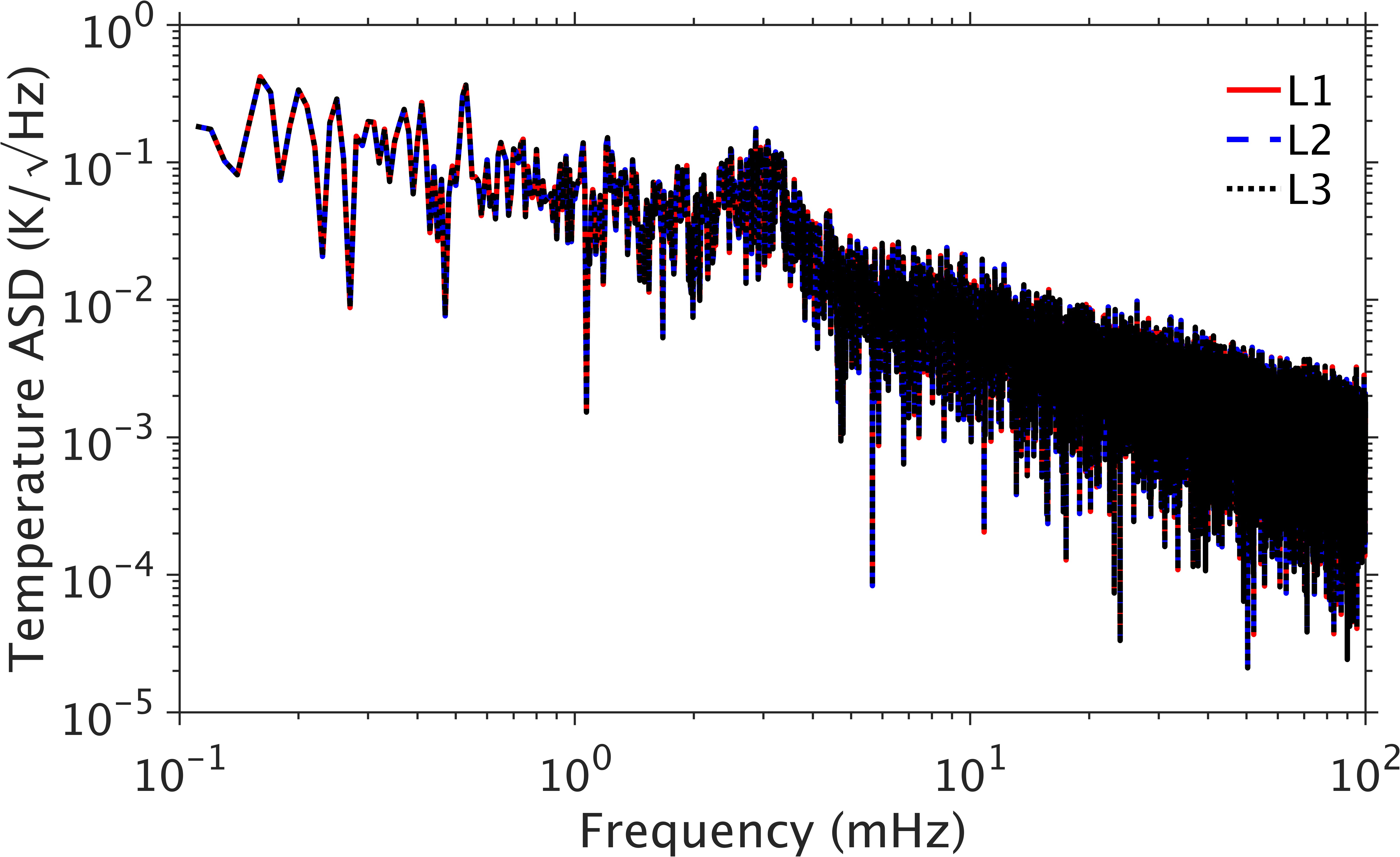}
		\caption{}
		\label{fig:temperature_asd}
	\end{subfigure}
	\vspace{0.01cm}
	\begin{subfigure}[h]{0.47\textwidth}
		\centering
		\includegraphics[width=\linewidth]{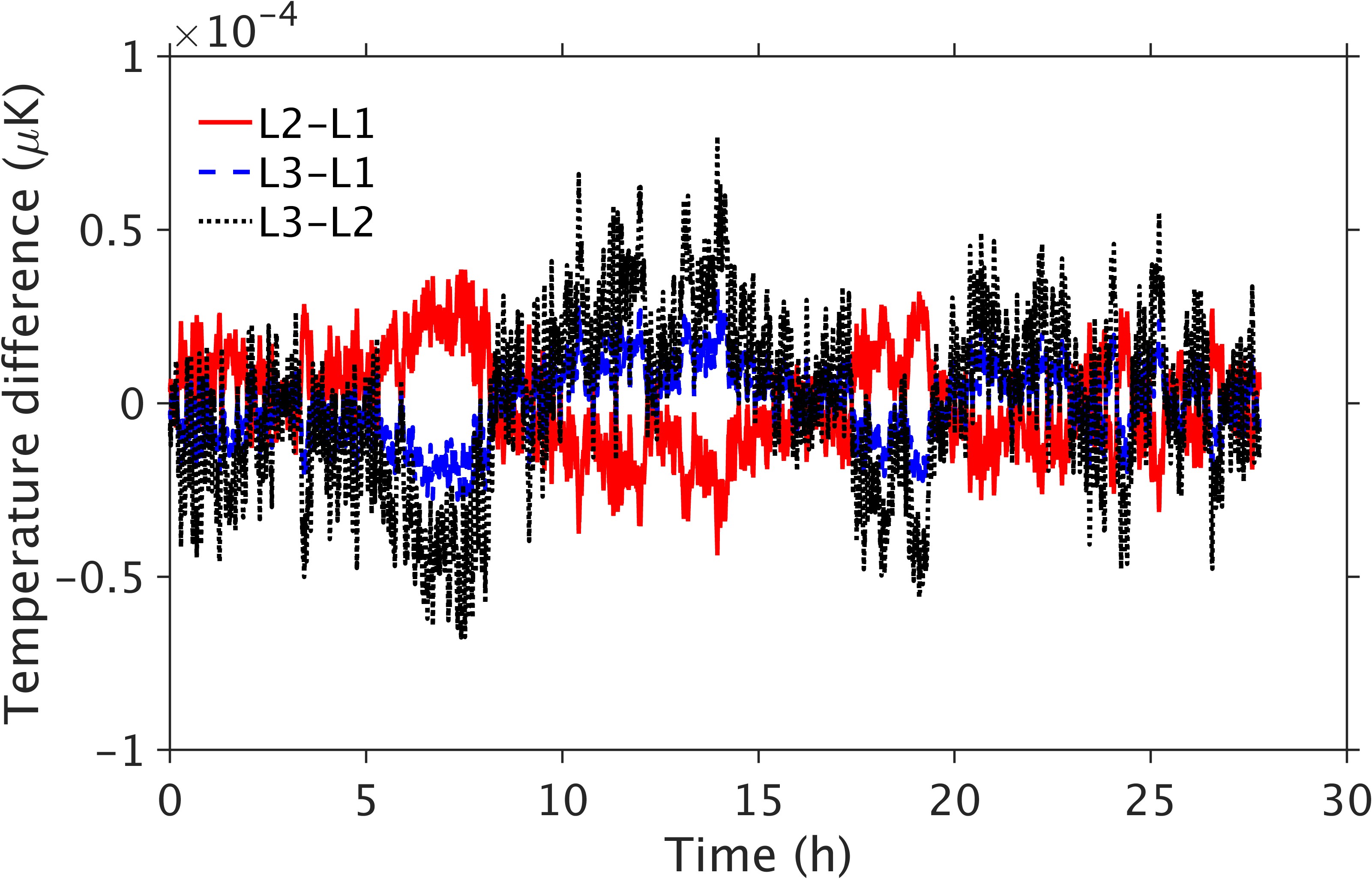}
		\caption{}
		\label{fig:temperature_differences}
	\end{subfigure}
	\caption{
		Temperature perturbations at the three PSE legs.
		(a) Surface temperature perturbations at L1, L2, and L3.
		(b) Amplitude spectral density of the temperature perturbations at the three legs.
		(c) Pairwise temperature differences between the PSE legs. The differences are of order $10^{-5}~\mu\mathrm{K}$, about $10^{-9}$ of the common-mode perturbation: the three legs, 0.61-0.63~m inside the shield edge, see practically the same temperature.
	}
	\label{fig:PSE_leg_temperature}
	
\end{figure*}

	\subsection{Frequency Range and Scientific Relevance}
	\label{sec:frequency_range}
	
	The model evaluates thermoelastic deformation driven by solar-flux variability between 0.01 and 100~mHz. This interval covers the low-frequency range relevant to long-period lunar seismology and overlaps the lower part of the frequency band proposed for lunar gravitational-wave instruments. In particular, it includes the region near $0.1~\mathrm{mHz}$ in which a lunar Slichter mode may be expected, depending on the adopted lunar interior model \cite{panningLILA}. The adopted solar-irradiance spectrum contains enhanced variability near 3.3~mHz, corresponding to the approximately five-minute solar oscillation band reported by \cite{frohlich2004solar}. This feature enters the model as an external solar-flux forcing and is propagated through the thermal and thermoelastic transfer functions, to determine whether it produces a corresponding feature in the temperature, displacement and tilt spectra. The resulting tilt spectrum represents an environmental-noise contribution that can be compared with independently predicted lunar-mode signals or instrumental noise requirements.

	Results are summarised in three bands: 0.01-0.05~mHz (the longest periods resolved), 0.08-0.12~mHz (Slichter-mode region) and 1-5~mHz (normal modes). These frequency bands are used only to identify where the calculated thermoelastic noise may affect low-frequency lunar measurements. Determining the detectability of a particular normal mode, Slichter-mode signal, or gravitational-wave signal requires a separate prediction of the signal amplitude and the instrumental response.
	
	\subsection{Superposed Vertical Displacement and Tilt Noise}
	\label{sec:tilt_noise}	
	The material properties and numerical parameters used in the calculation are given in Supplementary Section~\ref{app:numerical_parameters}. The observable instrument tilt is controlled by the differential vertical displacement of the three tripod legs.

	The temperature perturbation is almost identical at the three legs (Fig.~\ref{fig:PSE_leg_temperature}(a) and (b)). Its common-mode amplitude is $\sim0.01$~K, while the leg-to-leg differences are $\sim10^{-5}~\mu$K (Fig.~\ref{fig:PSE_leg_temperature}(c)). The differential displacement come from the elastic coupling described by Eq.~(\ref{eq:uz_feature}). Beneath an exactly centred tripod this displacement field is axisymmetric and the three legs move by the same amount. The surrounding instruments and the shiled centring error (Section~\ref{sec:apollo_geometry}) breaks this symmetry.
\begin{figure}[h!]
	\centering
	\includegraphics[width=0.9\textwidth,keepaspectratio]{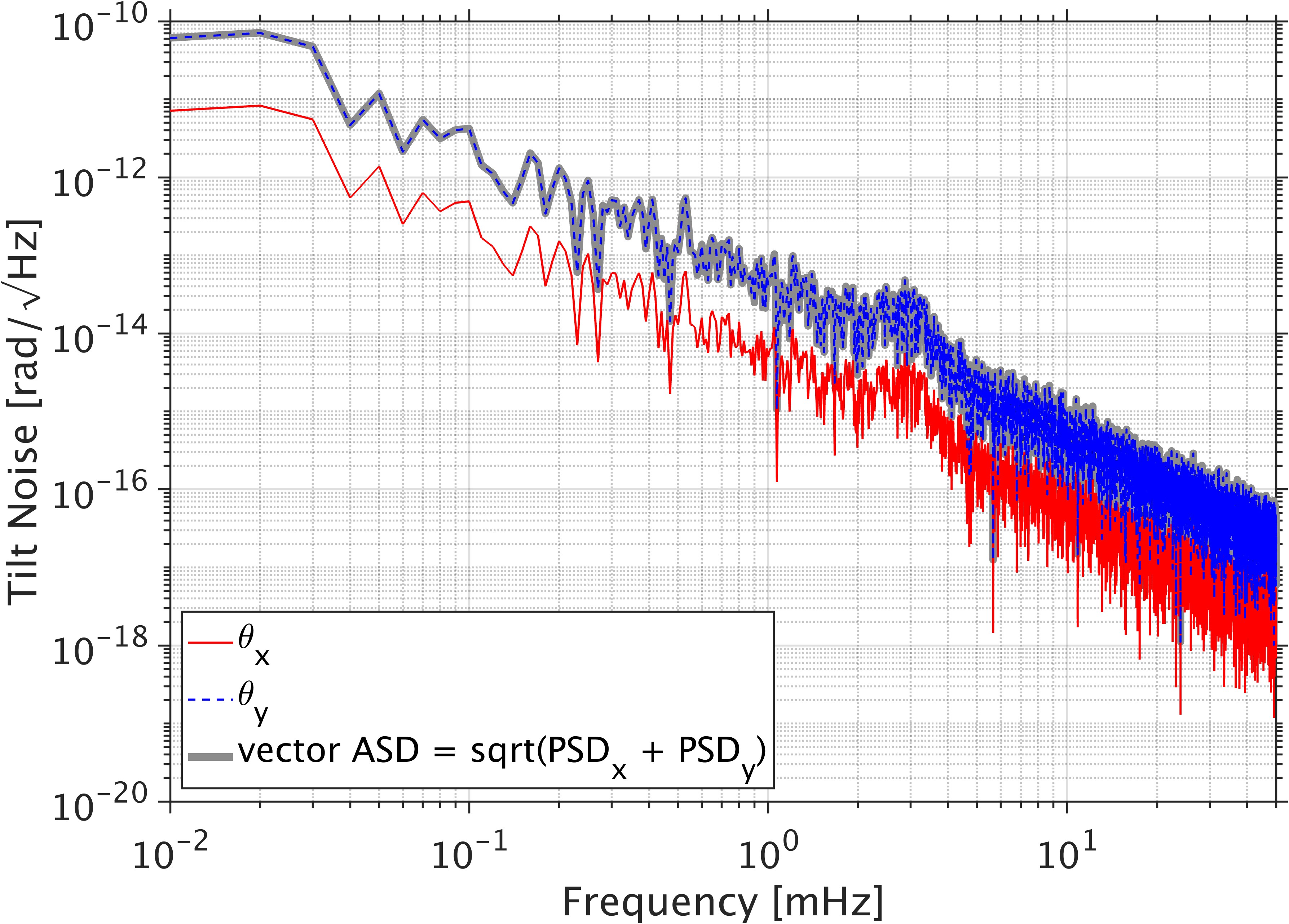}
	\captionsetup{font={small,stretch=0.8}}
	\caption{ Amplitude spectral density of thermoelastic tilt induced by the combined PSE thermal shield and surrounding Apollo 14 ALSEP instruments, computed with the elastic half-space displacement (Eq.~\ref{eq:uz_feature}). The tilt decreases steadily with frequency and is therefore mainly a long-period effect, with a secondary feature at the 3.3~mHz solar oscillation band. }
	\label{fig:tilt_psd}
\end{figure}
Thus the differential displacement between the three legs is $\approx7\times10^{-14}$~m rms (Fig.~\ref{fig:thermoelastic_displacement_tilt}(a) and (b)). The resulting tilt has an rms amplitude of $3.6\times10^{-13}$~rad, mostly in the component $\theta_y$ (Fig.~\ref{fig:thermoelastic_displacement_tilt}(c)). The conversion of leg displacements into tripod tilt is detailed in Supplementary Section~\ref{app:stochastic_thermal}.

	The corresponding frequency-domain representation is shown in Fig.~\ref{fig:tilt_psd}. The tilt ASD decreases steadily with frequency, by about one order of magnitude from one band to the next, so thermoelastic tilt is mainly a long-period effect; at low frequency the solar-flux variability is largest and the temperature perturbation penetrates deepest. The enhanced solar variability near 3.3~mHz \cite{frohlich2004solar} appears as a secondary feature but is not the maximum of the spectrum. The shield, through the centring error, contributes most of the tilt, while the surrounding ALSEP components contribute about an order of magnitude less. Because the tilt of every source is proportional to the same solar-flux spectrum, these ratios are the same in the three bands. For an exactly centred tripod, the shield contribution vanishes by symmetry and only the neighbouring instruments remain, giving a tilt about eight times smaller. These results assume an opaque shield; Section~\ref{sec:mli_apollo} shows that the partial transparency of the Apollo shield, combined with the shadow of the instrument body, produces a much larger tilt near local noon.


	\subsection{Partially transparent shield and the shadow of the instrument body}
	\label{sec:mli_apollo}
	
	Section~\ref{sec:tilt_noise} assumed that the shield blocks the sunlight completely. The Apollo shield, however, is an MLI blanket that lets a small fraction $\eta$ of the solar-flux fluctuation through to the ground (Section~\ref{sec:mli_model}). This brings a second feature into play, the shadow cast by the instrument body on the blanket (Fig.~\ref{fig:supp_setup}c,d). Although only a fraction $\eta$ of the fluctuation is involved, it can tilt the instrument as soon as the shadow is not centred on the tripod. By Eq.~(\ref{eq:three_zones}), the leaky shield is equivalent to an opaque shield weakened by $(1-\eta)$ plus the body shadow alone, of weight $\eta$. For the Apollo 14 shield ($R_s=0.75$~m, $\varepsilon_M=0.02$, $\eta=1/50$) and an instrument body of radius $R_b=0.14$~m on its own tripod, the displacement of leg $\ell$ is therefore
	\begin{equation}
		u_{z,\ell}=A_{\rm in}\left[(1-\eta)\,H_u(d^{(s)}_\ell;R_s)+\eta\,H^{(\eta h_r)}_u(d^{(b)}_\ell;R_b)\right]+\sum_{j}u_{z,j}(d_{\ell j}),
		\label{eq:uz_mli}
	\end{equation}
	where $H_u(d;R)$ is the displacement per unit flux contrast at a distance $d$ from the centre of a circular feature of radius $R$ (Eq.~\ref{eq:uz_feature}), $d^{(s)}_\ell$ and $d^{(b)}_\ell$ are the distances of the leg from the centres of the shield and of the body shadow, and the sum runs over the neighbouring instruments. 

	With the Sun at the zenith the body shadow is centred on the tripod, all legs lie at the same distance from its edge, and the body term produces no tilt. The tilt is then that of the opaque shield reduced by the factor $1-\eta$, so the transparency of the blanket alone does not increase the tilt. At the Apollo 14 site (latitude $3.64^\circ$S), however, the Sun remains $\zeta\approx2$-$5^\circ$ from the zenith at noon, because the sub-solar point stays within $\pm1.54^\circ$ of the equator, and the shadow of the 0.5-0.6~m tall body is displaced by $\Delta=1.75$-5.25~cm (Eq.~\ref{eq:shadow_shift}; Fig.~\ref{fig:supp_setup}b). The legs then lie at different distances from the shadow edge, and the body term controls the tilt. Fig.~\ref{fig:supp_compass} shows which leg comes out of the shadow for each direction of the shift. The tilt spectra are shown in Fig.~\ref{fig:mli_apollo} and their medians in the 1-5~mHz band in Table~\ref{tab:mli_tilt}, for shifts of 1, 2 and 5~cm; shifts of 2-5~cm bracket the noon range. For the nominal blanket ($\varepsilon_M=0.02$) the noon tilt is $\approx7$-$8\times10^{-11}~\mathrm{rad}/\sqrt{\mathrm{Hz}}$, more than three orders of magnitude above the opaque-shield value. 

	\begin{table}[h!]
		\centering
		\captionsetup{font={small,stretch=0.8}}
		\caption{Median tilt ASD in the 1-5~mHz band ($\mathrm{rad}/\sqrt{\mathrm{Hz}}$) of the model curves of Fig.~\ref{fig:mli_apollo}, for the body shadow shifted by 1, 2 and 5~cm, and ratio of the quietest 10\% of the Apollo 14 flat-mode noise near noon ($2.5\times10^{-9}~\mathrm{rad}/\sqrt{\mathrm{Hz}}$, Section~\ref{sec:apollo_noise}) to the modelled tilt for shifts of 2-5~cm, which bracket the noon range.}
		\label{tab:mli_tilt}
		\small
		\begin{tabular}{lcccc}
			\toprule
			Configuration & $\Delta=1$~cm & $\Delta=2$~cm & $\Delta=5$~cm & Apollo noise / tilt\\
			\midrule
			Opaque shield & \multicolumn{3}{c}{$2.2\times10^{-14}$} & $1.1\times10^{5}$\\
			MLI, $\varepsilon_M=0.02$ (nominal) & $3.6\times10^{-11}$ & $6.7\times10^{-11}$ & $8.3\times10^{-11}$ & 30-37\\
			MLI, $\varepsilon_M=0.1$ (degraded) & $1.8\times10^{-10}$ & $3.3\times10^{-10}$ & $4.1\times10^{-10}$ & 6-8\\
			MLI, $\varepsilon_M=0.2$ (degraded) & $3.4\times10^{-10}$ & $6.4\times10^{-10}$ & $7.9\times10^{-10}$ & 3-4\\
			No MLI (instrument body only) & $1.5\times10^{-9}$ & $2.7\times10^{-9}$ & $3.4\times10^{-9}$ & 0.7-0.9\\
			\bottomrule
		\end{tabular}
	\end{table}

	\subsection{Comparison with the Apollo 14 long-period noise near local noon}
	\label{sec:apollo_noise}
	To test whether the thermoelastic tilt modelled above could have limited the Apollo 14 observations, we measured the horizontal long-period noise of the Apollo 14 seismometer near local noon. We used the re-archived Apollo data \cite{nunn2022}, channels MH1 and MH2 of station S14, recorded in flat mode, which is the most sensitive mode at long periods and is available at S14 for 141 days in 1976. The records were corrected for the instrument response, cut into 3-hour segments, converted to acceleration spectra and expressed as an equivalent tilt $a/g_{\rm Moon}$. 
	\begin{figure}[h!]
		\centering
		\includegraphics[width=0.99\textwidth,keepaspectratio]{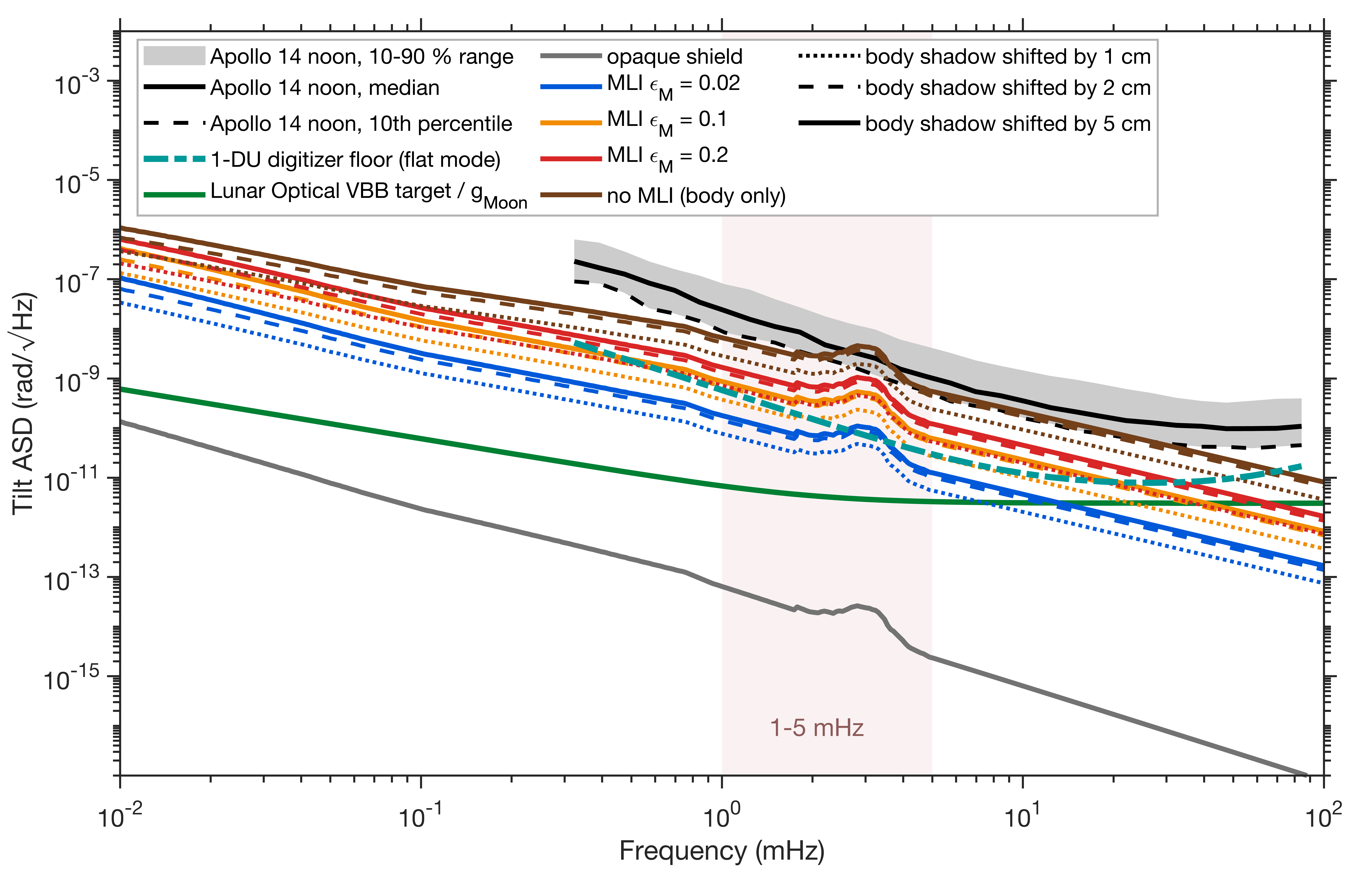}
		\captionsetup{font={small,stretch=0.8}}
		\caption{Modelled thermoelastic tilt ASD of the Apollo 14 PSE compared with the horizontal long-period noise of the Apollo 14 seismometer, expressed as an equivalent tilt $a/g_{\rm Moon}$. Data: flat-mode records of 1976, segments within one lunar hour of noon (10-90\% range, median and 10th percentile), and the 1-DU digitizer floor of the flat mode. Reference: tilt equivalent $A/g_{\rm Moon}$ of the Lunar Optical VBB design target \cite[Fig.~9]{branchesi2023}, extrapolated as $1/f$ below 0.2~mHz (Section~\ref{sec:design_requirement}). Model ($R_s=0.75$~m, 1~cm centring error, neighbouring instruments included): opaque shield (grey), and MLI shields of emissivity $\varepsilon_M=0.02$, 0.1 and 0.2 and no MLI (colours), with the body shadow shifted by 1 (dotted), 2 (dashed) and 5~cm (solid) towards $300^\circ$, the direction of the largest tilt, close to the southward noon shift. Shaded: 1-5~mHz.}
		\label{fig:mli_apollo}
	\end{figure}
	Near noon, the flat-mode noise in the 1-5~mHz band is $2.5\times10^{-9}~\mathrm{rad}/\sqrt{\mathrm{Hz}}$ for the quietest 10\% of the 52 retained segments and $4.8\times10^{-9}~\mathrm{rad}/\sqrt{\mathrm{Hz}}$ for their median, 20-40 times above the digitizer floor (Fig.~\ref{fig:mli_apollo}). 
	
	Fig.~\ref{fig:mli_apollo} and Table~\ref{tab:mli_tilt} compare these levels with the model. With an opaque shield the tilt is about $10^{5}$ times below the quietest Apollo noise near noon. With the nominal MLI and the noon shift of the body shadow (2-5~cm), the tilt rises to $\approx7$-$8\times10^{-11}~\mathrm{rad}/\sqrt{\mathrm{Hz}}$ in the 1-5~mHz band, but remains 30-37 times below that noise. A degraded blanket narrows this margin to 6-8 ($\varepsilon_M=0.1$) and 3-4 ($\varepsilon_M=0.2$), and the instrument body without any blanket reaches the Apollo noise level. The transparency of the shield is therefore a real source of tilt, which dominates the shield-related tilt as soon as the Sun is away from the zenith.

	The day-night contrast and the afternoon maximum of the observed noise point instead to thermal effects not included in this model, such as the thermal response of the sensor and of its mounting, thermoelastic tilts generated by thermal stresses on the cable, or large asymmetries in the thermal properties of the regolith, possibly caused by the astronauts' activity during the PSE installation. For future instruments, however, this tilt is clearly a significant source of noise. It exceeds the target of horizontal sensors aiming below $10^{-11}~\mathrm{rad}/\sqrt{\mathrm{Hz}}$, such as the Lunar Optical VBB (LOVBB), with a targeted sensitivity of $5\times10^{-12}~\mathrm{m\,s^{-2}}/\sqrt{\mathrm{Hz}}$ \cite{deRaucourt+2022,deraucourt2026} (green line in Fig.~\ref{fig:mli_apollo}), and even more that of the GW accelerometers proposed for the LGWA project \cite{Harms_2021,branchesi2023}, with targeted sensitivities below $5\times10^{-13}~\mathrm{m\,s^{-2}}/\sqrt{\mathrm{Hz}}$.

	\section{Implications for future lunar surface instruments}
	\label{sec:design}
	
	The analytical results provide direct guidance for the design and placement of surface-deployed lunar instruments that are mechanically coupled to the shallow regolith. This section examines how the size of the thermal shield and its emissivity control the tilt noise and the resulting design principles. 
	
	\subsection{Effect of shield radius on the tilt}
	\label{sec:design_radius}
\begin{figure}[h!]
		\centering
		\includegraphics[width=0.95\textwidth,keepaspectratio]{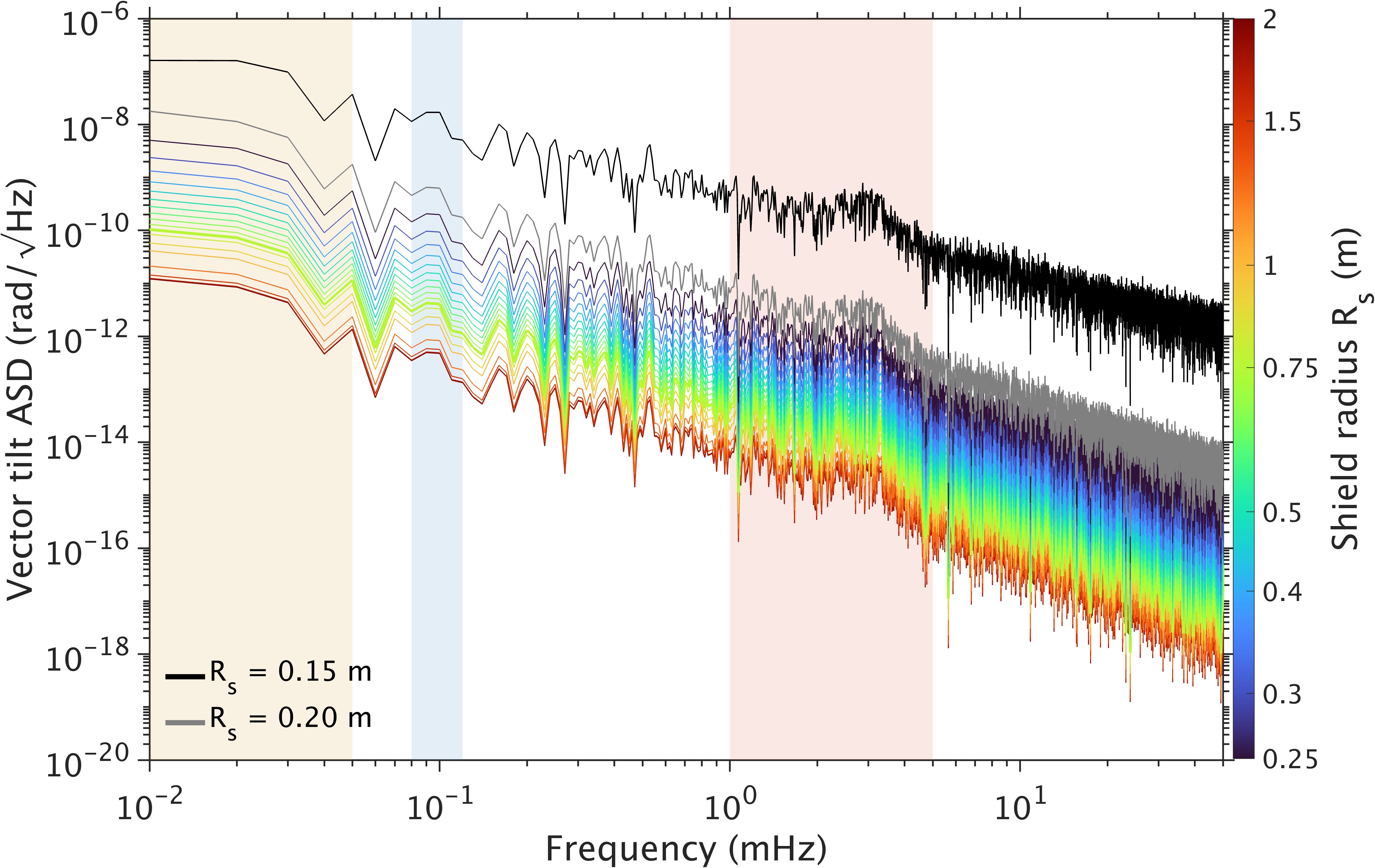}
		\captionsetup{font={small,stretch=0.8}}
		\caption{ Amplitude spectral density of thermoelastic tilt for shield radii ranging from 0.25 to 2 m in the Apollo 14 configuration (colour scale), and for the small shields $R_s=0.15$~m (black) and 0.20~m (grey), whose edge lies within a few centimetres of the legs. Shaded: 0.01-0.05~mHz, 0.08-0.12~mHz (Slichter mode) and 1-5~mHz (normal modes).}
		\label{fig:allradii}
	\end{figure}

	Fig.~\ref{fig:allradii} shows the tilt noise of the Apollo 14 layout for every shield radius, assuming first perfect shielding. All spectra decrease steadily with frequency and show the 3.3~mHz solar oscillation feature. Increasing the shield radius lowers the whole spectrum without changing its shape for $R_s\ge0.15$~m. . 
	The spectral shape results from the input solar-flux spectrum and the thermal response of the regolith. At low frequencies, radiative heat exchange dominates the surface temperature response, whereas at higher frequencies, conduction becomes increasingly important. 
	The thermal skin depth, $\delta=\sqrt{\kappa/(\pi\rho C_p f)}$ ,  decreases with frequency, confining temperature perturbations to progressively shallower depths. At a depth of (4.6$\delta$), the perturbation amplitude falls to approximately 1\% of its surface value. As summarized in Table~\ref{tab:depth}, the maximum penetration depth is approximately 8.4 cm at 0.01 mHz and decreases to 0.8 cm at 1 mHz. This shallow thermal penetration, together with the decreasing solar-flux variability at higher frequencies, explains why thermoelastic tilt is predominantly a long-period effect.
	
	\begin{figure}[h!]
		\centering
		\includegraphics[width=0.8\textwidth,keepaspectratio]{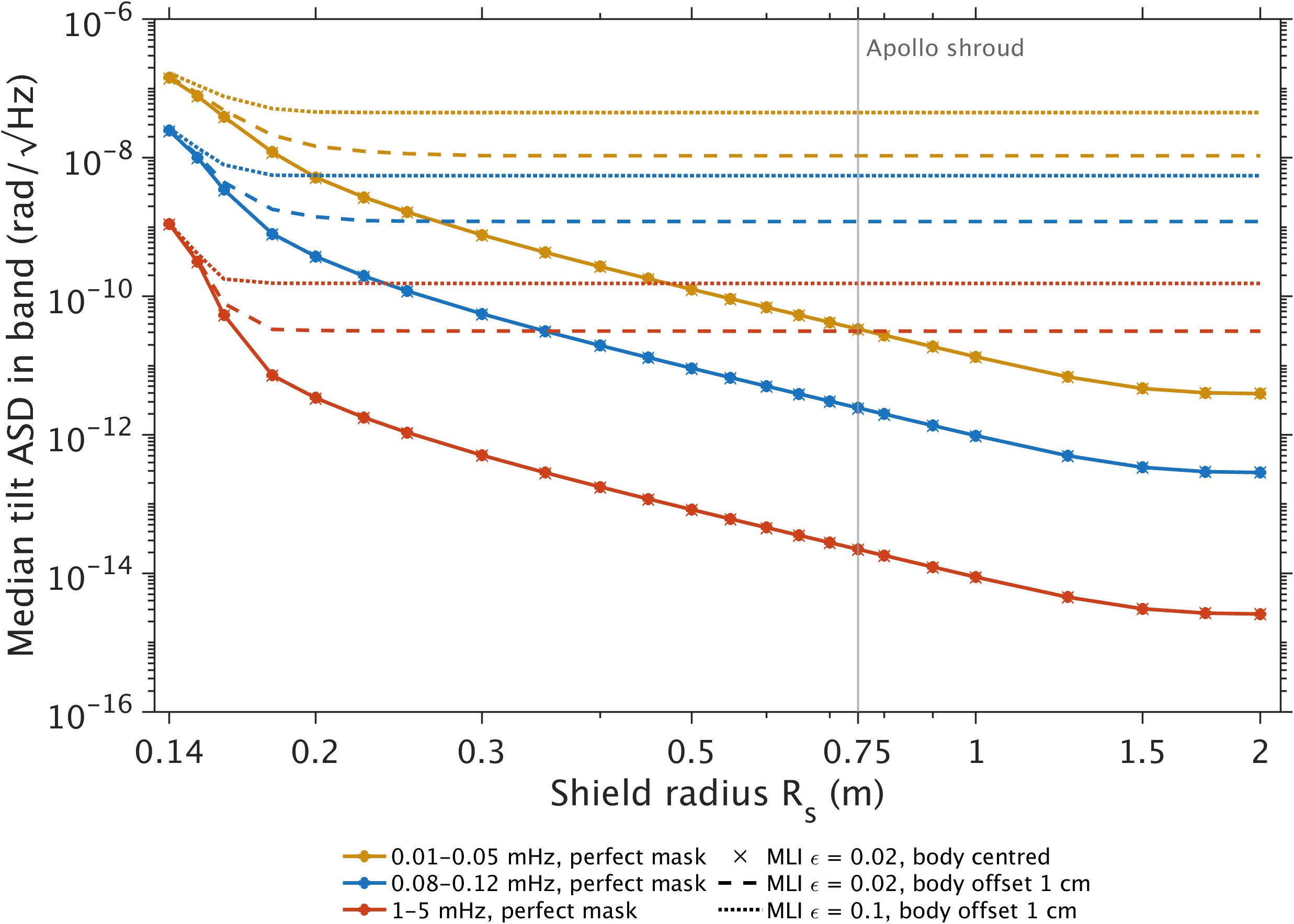}
		\captionsetup{font={small,stretch=0.8}}
		\caption{ Effect of shield radius $R_s$ on the thermoelastic tilt of the Apollo 14 layout, for $R_s=0.14$-2~m; $R_s=0.14$~m is the radius of the instrument body, i.e. no shield. Each point is the median of the vector tilt ASD within a frequency band at the Apollo 14 legs. Solid lines with dots: opaque shield with the 1~cm centring error. Crosses: MLI shield ($\varepsilon_M=0.02$) with the body shadow centred on the tripod, which coincides with the opaque shield ($\times0.98$). Dashed and dotted: MLI shield with the body shadow shifted by 1~cm, $\varepsilon_M=0.02$ and 0.1 (Section~\ref{sec:mli_apollo}). }
		\label{fig:Rchange}
	\end{figure}

	Fig.~\ref{fig:Rchange} shows three regimes, set by where the legs stand relative to the edge. For $R_s\lesssim0.2$~m the shield edge comes within a few centimetres of the legs (L2 and L3 are only 4~mm inside it for $R_s=0.14$~m), the legs lie within the thermal band at the edge, so they sit on regolith at different temperatures, and the tilt rises steeply, by a factor of about 90-1000 between $R_s=0.25$ and 0.14~m. For $0.25\lesssim R_s\lesssim1.5$~m the legs are at least six skin depths inside the edge and see the same temperature. The tilt then comes only from the gradient of the displacement field radiating from the edge. For a centred tripod this field is axisymmetric and the gradient cancels at the three legs; here the 1~cm centring error (Section~\ref{sec:apollo_geometry}) leaves a small residual, which weakens as the edge moves further away. Above $R_s\approx1.5$~m the tilt is dominated by the four neighbouring instruments. Each blocks sunlight over its own footprint and produces its own displacement field, which decreases with distance from it. The seismometer lies 2.3-4.2~m to one side of each instrument, so this field slopes across the tripod and tilts it.

	For an MLI shield the picture changes (dashed and dotted lines in Fig.~\ref{fig:Rchange}). Once the body shadow is shifted, the tilt is controlled by the edge of the shadow, which lies close to the legs whatever the shield radius, and it no longer decreases with $R_s$ above $\approx0.25$~m. Enlarging the shield then does not help; only a lower transmission of the blanket or a smaller shift of the shadow reduces the tilt. The point at $R_s=0.14$~m of the opaque-shield curve applies the 1~cm centring error to a shield of the size of the instrument body; a separate small shield could be off-centre in this way, but the shadow of the body itself is centred on its tripod when the Sun is at the zenith. 

	\begin{table}[h!]
		\centering
		\caption{Thermal penetration depth in each band. $\delta$: depth at which the temperature perturbation has decreased to $1/e$ (about 37\%) of its surface value. 1\% depth: depth at which it has decreased to 1\% of its surface value, given for the lowest frequency of each band, where it is largest.}
		\label{tab:depth}
		\small
		\begin{tabular}{lcc}
			\hline
			\textbf{Frequency Band} & \textbf{Skin depth} $\delta$ & \textbf{1\% depth} \\
			\hline
			0.01-0.05 mHz & 8.1-18 mm & 8.4 cm \\
			0.08-0.12 mHz (Slichter mode) & 5.3-6.4 mm & 3.0 cm \\
			1-5 mHz (normal modes) & 0.81-1.8 mm & 0.84 cm \\
			\hline
		\end{tabular}
	\end{table}

	\subsection{Comparison with seismometer self-noise and faint geophysical signals}
	\label{sec:design_requirement}
A tilt $\theta$ projects lunar gravity on the horizontal components of a seismometer as $g_{\rm Moon}\theta$ \cite{rodgers1968,murdoch2017wind}, so an instrument with acceleration self-noise $A$ is affected by thermoelastic tilt when the tilt ASD approaches $A/g_{\rm Moon}$ \cite{murdoch2017wind}. This first-order coupling applies to a horizontal sensitive axis. For a vertical axis, tilt changes the projection of gravity only at second order, $g_{\rm Moon}(1-\cos\theta)\approx g_{\rm Moon}\theta^2/2$, and otherwise enters through the cross-axis and mechanical response of the instrument. The values $A/g_{\rm Moon}$ are therefore the tilt noise that a horizontal sensor with self-noise $A$ could tolerate, that are used here as reference levels. 	 Table~\ref{tab:instruments} converts the self-noise of lunar seismometers into $A/g_{\rm Moon}$ and gives the band in which each value is stated. 
	\begin{table}[h!]
		\centering
		\caption{Self-noise of lunar seismometers, the overlap of the frequency at which it is stated with the frequency range of this study (0.01-100~mHz), and the equivalent tilt noise $A/g_{\rm Moon}$ for a horizontal sensitive axis. The frequencies are given in the text below.}
		\label{tab:instruments}
		\footnotesize
		\begin{tabular}{>{\raggedright\arraybackslash}p{0.38\linewidth}ccc}
			\hline
			\textbf{Lunar Instrument} & \shortstack{\textbf{Self-noise \textit{A}} \\($\mathrm{m\,s^{-2}}/\sqrt{\mathrm{Hz}}$)} & \shortstack{\textbf{Overlap}\\(mHz)} & \shortstack{\\\textbf{$A/g_{Moon}$}\\($\mathrm{rad}/\sqrt{\mathrm{Hz}}$)} \\
			\hline
			LGN VBB (proposed) \cite{haviland2022} & $3.5\times10^{-11}$ & 10-100 & $2.2\times10^{-11}$ \\
			Lunar Optical VBB target \cite{branchesi2023} & $5\times10^{-12}$ & 10-100 & $3.1\times10^{-12}$ \\
			LGWA (proposed) \cite{ajith2025lgwa} & $3.9\times10^{-13}$ & 100 & $2.4\times10^{-13}$ \\
			\hline
		\end{tabular}
	\end{table}

	 For the Lunar Optical Very Broad Band seismometer under development at IPGP \cite{deraucourt2026}, Branchesi et al.~\cite{branchesi2023} give a design target of $5\times10^{-12}~\mathrm{m\,s^{-2}}/\sqrt{\mathrm{Hz}}$ between 0.01 and 1~Hz, equivalent to $A/g_{\rm Moon}$ of $3.1\times10^{-12}~\mathrm{rad}/\sqrt{\mathrm{Hz}}$. The sensitivity at 1~mHz is about $1.1\times10^{-11}~\mathrm{m\,s^{-2}}/\sqrt{\mathrm{Hz}}$, or about $6.8\times10^{-12}~\mathrm{rad}/\sqrt{\mathrm{Hz}}$; below 1~mHz it degrades roughly as $1/f$ down to 0.2~mHz, the lowest frequency of the published curve, and is extrapolated below as $1/f$. This tilt equivalent is shown in Fig.~\ref{fig:mli_apollo}. As discussed above, $A/g_{\rm Moon}$ is the level relevant to a horizontal sensor with the Lunar Optical VBB self-noise; the tilt requirement of the instrument itself depends on its tilt-to-output response. For an opaque shield the modeled tilt remains below this level in all bands.

	These conclusions hold however only for an opaque shield. For an MLI shield the shifted shadow of the instrument body dominates (Section~\ref{sec:mli_apollo}), and with the Apollo values ($\varepsilon_M=0.02$, $R_s=0.75$~m, noon shift $\Delta=2$-5~cm) the median tilt ASD exceeds the median tilt equivalent of the Lunar Optical VBB target (Fig.~\ref{fig:mli_apollo}) by factors of 16-20 in the 1-5~mHz band, corresponding to the low angular order normal modes, 40-53 in the 0.08-0.12~mHz band, corresponding to the Slichter mode band, and 77-118 in the 0.01-0.05~mHz band, independently of the shield radius. Even a 1~cm shift gives factors of 9, 21 and 41. Since the shifted disk is expected to overestimate the tilt of a real shadow (Section~\ref{sec:mli_model}), these factors are upper estimates, but they show that for a seismometer of this class the optical transmission of the shield and the symmetry of the instrument's own shadow about its feet are design parameters as important as the shield size. Future instruments aiming to reach horizontal noise much below $10^{-10}~\mathrm{m\,s^{-2}}/\sqrt{\mathrm{Hz}}$ will not only request meter radius shield, but also much better MLI effective emissivity, possibly below 0.001, achieved not only by multi-layering but also by a minimization of the contact points between the shield and the ground. An alternative might also consist in thermal shield with significant thermal masses, which might then be able to filter the sun-driven fluctuations. This however will likely request much larger masses than those requested for simple multi-layers MLI structures. 
    
    Such an effort will also be requested for Gravity wave detectors in PSR based on tilt-sensitive accelerometers, like LGWA, as only a factor 10 of noise reduction might be obtained in PSR. This shows also that even on the Moon, sensors affected by tilts to the second order, like vertical seismometers or like the LILA strainmeters \cite{jani2025laserinterferometerlunarantenna,Panning2026}, will have much better signal to noise ratio. 
	
		\subsection{Clearance from other surface features}
\label{sec:design_clearance} 
\begin{figure}[h!]
	\centering
	\includegraphics[width=0.9\textwidth,keepaspectratio]{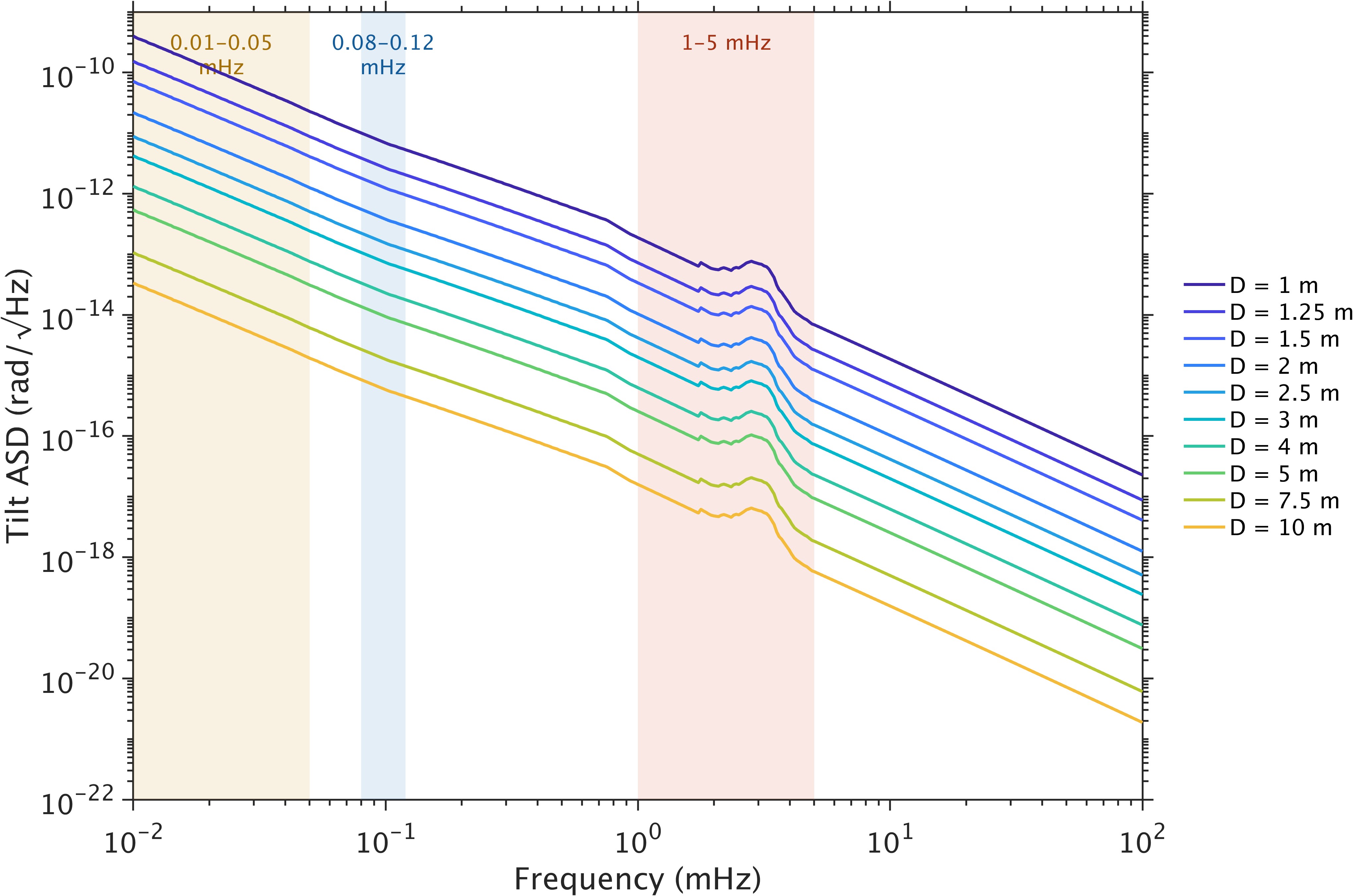}
	\captionsetup{font={small,stretch=0.8}}
	\caption{Thermoelastic tilt ASD from an opaque PSE shield ($R_s=0.75$~m) and one surface feature of radius 0.25~m, centred at distance $D$ from the shield centre along $-y$, at the legs of the Apollo 14 tripod centred on the shield (no neighbouring instruments). For the centred tripod the opaque shield contributes no tilt. Shaded: 0.01-0.05~mHz, 0.08-0.12~mHz and 1-5~mHz.}
	\label{fig:patch}
\end{figure}

Other surface features near the seismometer, such as hardware or rocks, also produce tilt. To estimate how far away they must be, we added to the PSE shield ($R_s=0.75$~m; no ALSEP instruments) a single circular feature with albedo of 1, centred at a distance $D=1$-10~m from the shield centre, so that it lies entirely outside the shield. Here the Apollo 14 tripod is taken as centred on the shield, the shield is opaque, so the shield itself produces no tilt and the tilt comes from the feature alone. Fig.~\ref{fig:patch} shows the tilt ASD for $D=1$-10~m. The spectral shape is the same at every distance, and for $D\gtrsim2$~m the tilt decreases approximately as $D^{-4}$; in the 0.01-0.05~mHz band its median ASD falls from $1.0\times10^{-10}$ at 1~m to $8.4\times10^{-15}~\mathrm{rad}/\sqrt{\mathrm{Hz}}$ at 10~m. 

Taken together, these findings indicate that minimizing thermoelastic tilt is not simply a matter of increasing the thermal shield size. For the Apollo 14 configuration with an opaque shield, enlarging the shield beyond approximately 1.5 m offers limited improvement because nearby surface features become the dominant contributors. Accurate centering of the instrument and sufficient clearance from surrounding hardware are therefore equally important considerations. If the shield transmits part of the solar flux, as an MLI blanket does, the shadow of the instrument body becomes the main source of tilt whenever the Sun is away from the zenith, and a low transmission of the blanket, or a shading geometry that keeps the shadow symmetric about the feet, matters more than the shield radius. More generally, the analytical framework provides a means of translating instrument sensitivity requirements into constraints on shield dimensions, deployment geometry, and separation distances, offering a practical basis for mitigating thermally induced tilt in future lunar seismic experiments.
	
	\section{Limitations and Applicability}
	\label{sec:limitations}
	
	The analytical framework developed in this study is designed to isolate the thermal and
	thermoelastic effects arising from localized surface heterogeneities under controlled and
	physically transparent assumptions. While this approach provides valuable insight into the
	mechanisms governing lateral heat diffusion and instrument tilt, several limitations should
	be acknowledged when applying the results to real lunar deployment scenarios.
	\begin{enumerate}
		\item The regolith is modeled as a laterally homogeneous, isotropic, semi-infinite elastic
		half-space with constant thermal and mechanical properties. In reality, lunar regolith
		properties may vary with depth due to compaction, layering, or compositional changes, and
		may exhibit local heterogeneity at meter scales. Such variations could modify the detailed
		amplitude and phase of the thermal response but are not expected to alter the fundamental
		mechanism by which localized surface features generate lateral temperature gradients and
		thermoelastic deformation.
		\item The surface energy-balance boundary condition is linearized about a mean equilibrium
		temperature and evaluated using harmonic forcing. This approximation is appropriate for
		analyzing diurnal and long-period thermal perturbations but neglects strongly nonlinear
		effects that may arise during rapid transients, eclipse events, or extreme temperature
		excursions. Similarly, radiative heat exchange between nearby surface features is not
		explicitly modeled and may influence the thermal field in complex lander environments.
				\item The thermoelastic coupling is computed for a homogeneous, isotropic elastic half-space with a traction-free surface. The increase of the regolith stiffness with depth and its layering are not included, and lateral compliance, slip, or plastic deformation at the foot-regolith interface are neglected. While these assumptions are reasonable for small thermally induced strains, more complex contact mechanics could become relevant for loosely coupled instruments or in regions with unconsolidated surface material.
		\item The seismometer is idealized as a rigid tripod that samples deformation at discrete
		points. This representation captures the essential geometric filtering of near-surface
		deformation but does not account for instrument-specific structural compliance or internal
		mechanical resonances. As a result, the predicted tilt should be interpreted as a lower-order
		estimate of the thermoelastic contribution rather than a complete instrument response model.
        \item We have here not considered the noise generated by Earth shine  \cite{Glenar2019}, which is about 0.01\% of the sun radiance and might therefore correspond to a 40K radiance black-body. Very large hourly fluctuation have been reported, of about 20\% \cite{Langford2009,MontasRodrguez2007}, which therefore might generate temperature variation of ${T_0 \over 4}{\delta\Phi_{earthshine} \over \Phi_{Sun}}$ and of the order of mK over hours. 
		\item The regolith is represented by a single layer with the density of the loose surface (fluff) layer, $\rho=1200~\mathrm{kg\,m^{-3}}$. Because the temperature perturbations penetrate less than $\approx8$~cm at the frequencies considered (Table~\ref{tab:depth}), the response is controlled by this surface layer; the increase of density with depth is not included. A density of $1800~\mathrm{kg\,m^{-3}}$ lowers all tilts by a factor of 1.5-1.75 and leaves the conclusions unchanged.
		\item The shadow of the instrument body on an MLI shield is represented by a circular disk shifted by $\Delta$. The real shadow is elongated, with the ground beneath the body remaining dark, and the lateral conduction of heat along the metallized layers of the blanket smooths the contrast at its edge; both are expected to reduce the tilt. The geometry is also frozen: the Sun moves by about $0.5^\circ$ per hour, so the shadow slides by centimetres over the periods of the 0.01-0.05~mHz band, and its slow motion is itself a large forcing outside the small-perturbation model. The reduced radiative coefficient beneath the shield is applied to the body term only, and the shadows of the neighbouring instruments are not shifted with the Sun.
		\item The Apollo 14 noise near noon is measured from the flat-mode records of 1976 only, about five lunations; two thirds of the noon segments are rejected, which biases the noon statistics low, and the instrument responses are used as archived. The difference between the flat and peaked recording modes is not resolved.
	\end{enumerate}
	Despite these limitations, the analytical formulation is particularly well suited for parametric exploration of shielding
	geometry, surface heterogeneity, and deployment configuration, and can serve as a
	computationally efficient complement to fully numerical thermal-mechanical simulations. More generally, the approach may be extended to other airless bodies with low thermal
	diffusivity and strong radiative forcing, such as asteroids or the surfaces of Mercury and
	icy satellites, provided that appropriate material properties and forcing timescales are
	used.

	\section{Summary}
	
	This study developed an analytical framework to evaluate thermoelastic tilt noise produced by localized thermal heterogeneity around lunar surface seismometers. Thermal shields and nearby surface hardware modify the local absorption of solar radiation, creating lateral temperature gradients in the shallow regolith. These gradients can drive differential vertical deformation across the seismometer legs and appear as long-period tilt noise. Using separation of variables and Hankel transforms, we derived a solution for the temperature perturbation beneath circular surface features and coupled it to thermoelastic displacement. Because the governing equations are linear, the effects of multiple nearby instruments can be superposed. \\
	Applied to the Apollo 14 Passive Seismic Experiment, the model shows that the tilt comes from the elastic coupling of the regolith. For an opaque shield, a perfectly centred tripod would not tilt at all; the small centring error of a manually deployed instrument breaks this symmetry, and the neighbouring instruments add a contribution about an order of magnitude smaller. The resulting tilt is mainly a long-period effect, because the solar flux varies most, and heat penetrates deepest, at low frequencies.\\
	The Apollo shield, however, is a multilayer insulation blanket that lets about 1/50 of the solar-flux fluctuation through to the ground. This brings into play the shadow of the instrument body on the blanket, whose edge lies very close to the legs. With the Sun exactly overhead the shadow is centred on the tripod and causes no tilt, but at the Apollo 14 site the noon Sun is a few degrees from the zenith, the shadow moves by a few centimetres, and one leg comes out into the leaked light. This raises the tilt by about three orders of magnitude compared with an opaque shield. The tilt then depends on how much light the blanket transmits and on the shift of the shadow, but no longer on the size of the shield. The results show that shield transmission, albedo contrast, shield size, and instrument placement all control the magnitude of thermoelastic tilt.\\
	The same tilt modelled in this study, nevertheless exceeds the noise target of a next-generation very broad band seismometer such as the Lunar Optical VBB, and even more that of the tilt-sensitive accelerometers proposed for lunar gravitational-wave detection. Reducing it requires more than a larger shield. The instrument should be accurately centred on its shield, kept away from other hardware and rocks, and covered by a blanket of very low transmission, with a shading geometry that keeps its own shadow symmetric about its feet; a shield with a large thermal mass is a possible alternative. Sensors that are sensitive to tilt only at second order, such as vertical seismometers or strainmeters, are much less affected. More generally, the main design goal should be to minimize lateral thermal gradients across the instrument footprint rather than only the absolute temperature variations, and the framework presented here provides an efficient way to turn instrument requirements into constraints on shield size and transmission, deployment geometry and separation distances.
	
%

	\clearpage
	\bibliographystyle{unsrt}
	\bibliography{agusample_v5.bib}

\end{document}